\documentclass{imammbpreprint}

\jno{dqnxxx}

\usepackage{natbib}

\usepackage{graphicx}
\usepackage{amsmath,amsthm,bm,mathrsfs}

\renewenvironment{proof}[1][Proof]{\noindent\textit{#1. } }{\hfill$\square$}

 \newtheoremstyle{theorem}{6pt}{6pt}{\rm}{}{\sffamily}{ }{ }{}
 \theoremstyle{theorem}

 \newtheoremstyle{algorithm}{6pt}{6pt}{\rm}{}{\sffamily}{ }{ }{}
 \theoremstyle{algorithm}

 \newtheoremstyle{lemma}{6pt}{6pt}{\rm}{}{\sffamily}{ }{ }{}
 \theoremstyle{lemma}

\newtheoremstyle{case}{6pt}{6pt}{\rm}{}{\sffamily}{. }{ }{}
 \theoremstyle{case}

 \newtheoremstyle{statement}{6pt}{6pt}{\rm}{}{\sffamily}{ }{ }{}
\theoremstyle{statement}

 \newtheoremstyle{corollary}{6pt}{6pt}{\rm}{}{\sffamily}{ }{ }{}
 \theoremstyle{corollary}

  \newtheoremstyle{definition}{6pt}{6pt}{\rm}{}{\sffamily}{ }{ }{}
 \theoremstyle{definition}

\newtheoremstyle{example}{6pt}{6pt}{\rm}{}{\sffamily}{ }{ }{}
\theoremstyle{example}

\newtheoremstyle{remark}{6pt}{6pt}{\rm}{}{\sffamily}{ }{ }{}
\theoremstyle{remark}

\newtheoremstyle{approximation}{6pt}{6pt}{\rm}{}{\sffamily}{ }{ }{}
\theoremstyle{approximation}

\newtheoremstyle{scheme}{6pt}{6pt}{\rm}{}{\sffamily}{ }{ }{}
\theoremstyle{scheme}

\newtheoremstyle{Algorithm}{6pt}{6pt}{\rm}{}{\sffamily}{ }{ }{}
\theoremstyle{Algorithm}

\newtheoremstyle{Assumption}{6pt}{6pt}{\rm}{}{\sffamily}{ }{ }{}
\theoremstyle{Assumption}

\newtheoremstyle{proposition}{6pt}{6pt}{\rm}{}{\sffamily}{ }{ }{}
\theoremstyle{proposition}
\newtheorem{proposition}{\sc Proposition}[section]

\newtheoremstyle{hypo}{6pt}{6pt}{\rm}{}{\sffamily}{ }{ }{}
 \theoremstyle{hypo}

  \newtheoremstyle{Step}{6pt}{6pt}{\rm}{}{}{ }{ }{}
 \theoremstyle{Step}

\newcommand{\ds}{\displaystyle}
\newcommand{\vs}{{\bf v_S}}
\newcommand{\vl}{{\bf v_L}}
\newcommand{\vu}{{\bf v}}
\newcommand{\T}{{\bf T}}
\newcommand{\m}{{\bf m}}
\newcommand{\one}{{\bf I}}
\newcommand{\eps}{\varepsilon}
\newcommand{\pa}{\partial}

\numberwithin{equation}{section}

\begin{document}

\title{A fluid dynamics multidimensional model of biofilm growth: stability, influence of environment and sensitivity}
\author{{\sc F. Clarelli$^1$, C. Di Russo$^2$, R. Natalini$^1$ \& M. Ribot$^3$}
\\[2pt]
$^1$ Istituto per le Applicazioni del Calcolo ``M. Picone'', Consiglio Nazionale delle Ricerche, Italy\\ [6pt]
$^2$ Laboratoire MAPMO (UMR CNRS 7349), Universit\'e d'Orl\'eans, 
F\'ed\'eration Denis Poisson, F-45067 Orl\'eans cedex 2,  France\\ [6pt]
$^3$ Laboratoire J. A. Dieudonn\'e, UMR CNRS 7351, Universit\'e de Nice-Sophia Antipolis, Parc Valrose, F-06108 Nice Cedex 02, France \&  Project Team COFFEE, INRIA Sophia Antipolis, France\\ [6pt]
\vspace*{6pt}}
\pagestyle{headings}
\markboth{F. CLARELLI, C. DI RUSSO, R. NATALINI, M. RIBOT}{\rm   MODEL OF BIOFILM GROWTH :  STABILITY, INFLUENCE OF ENVIRONMENT AND SENSITIVITY}
\maketitle


\begin{abstract}
{In this article, we study in  details the fluid dynamics  system proposed in \cite{cdnr}  to model the formation of  cyanobacteria biofilms.  After  analyzing  the linear stability of the unique non trivial equilibrium of the system, we introduce in the model the influence of light and temperature, which are two important factors  for the development of cyanobacteria biofilm. Since the values of the coefficients we use for our simulations  are estimated through information found in the literature, some sensitivity and robustness analyses on these parameters are performed. 
All these elements enable us to control and to validate the model we have already derived and to present some numerical simulations in the 2D and the 3D cases. }
{Fluid dynamics model, Hyperbolic equations,  Phototrophic biofilms, Sensitivity, Stability.}
\end{abstract}

\section{Introduction}
\label{intro}

If the problem of chemical degradation has been for many decades the main concern for conservation and restoration studies, there is now an increasing experimental evidence that biodegradation phenomena have also to be taken into account, since 
a large part of the deterioration of monumental artifacts is due to biological factors, and more specifically, in connection with biofilm structures. From the mathematical point of view, there is a long series of models, which have been proposed in the last 30 years to describe the evolution of biofilms, but these studies were mainly addressed to specific situations concerning biomedical issues, petroleum extraction  or sewage treatments, see for instance \cite{Delf} or \cite{kla09}. 

Here we are interested in particular in the formation and evolution of biofilms, with a special regard to their development on fountains walls, i.e.: on stone substrates and under a water layer. This kind of biofilms causes many damages, such as unaesthetic biological patinas, decohesion and loss of substrate material from the surface of monuments or degradation of the internal structure. Let us recall that a biofilm is a complex aggregation of various microorganisms like bacteria, cyanobacteria, algae, protozoa and fungi, all embedded in an extracellular matrix of polymeric substances, usually indicated as EPS. The EPS acts as a barrier which enhances resistance to antibiotics, to the immune response, to disinfectants or cleaning fluids. Typically a biofilm contains water, but it can be considered in a solid/gel phase. Biofilms can develop on surfaces which are in permanent contact with water, i.e. solid/liquid interfaces, but more general behaviours are also quite common. 

To describe the growth of a biofilm structure, in \cite{cdnr} we introduced a continuous fluid dynamics model. This model was built in the framework of mixture theory, see  \cite{raj95} or \cite{astanin}, and we conserved the finite speed of propagation of the fronts. In practice, we started  from some   balance equations for mass and momentum conservation, and added some physical constraints  and assumptions about the behaviour of the biological aggregate and its interaction with the surrounding liquid. In the present paper, we consider again this model and we give some more precise arguments to support it and to a specify its behaviour. 

More precisely, after a short presentation of the model, a first part of the paper is devoted to assess the linear stability of the model in the one dimensional case, around the unique non trivial equilibrium of the system. To obtain this result, we consider the case where nutrients and temperature are not limiting factors and   do not play a significant role; however, it would be simple to add their dependence by perturbation arguments. 
In the second part of the paper, we  estimate the coefficients and their dependence  on  light and temperature, the latter being not included in our previous paper, so to establish the behaviour of the model under different physically relevant conditions.

Next, since our model contains a large number of parameters, we study the sensitivity and the robustness of them, with respect to some biological reference values. Our analysis reveals the crucial dependence of our model on some specific parameters, i.e.: the bacteria growth rate and the stress tensor coefficient. 

Finally, once  the model calibrated, we simulate its evolution under different conditions in 2D and 3D,  observing the influence of light and temperature on the biofilm growth.

\section{The fluid dynamics model}

Let us first recall the model we consider in this article. 
To describe the complex process of biofilm  growth, we use the model presented in  \cite{cdnr}. We define   four different phases, which are: cyanobacteria ($B$), dead cells ($D$), EPS ($E$), and liquid ($L$). Let the concentration of biomass be $C_{\phi} = \rho_{\phi} \phi$, where $\phi = B,D,E,L$ is the volume fraction of each component and 
$\rho_{\phi}$ is the mass density ($[g/cm^3]$) of $\phi$. We assume biomasses as incompressible and Newtonian, thus $\rho_{B}$, $\rho_{E}$, $\rho_{D}$ and $\rho_{L}$ are positive constants. For simplicity, in this first approach, we also assume that the phases have all the same density. Thus, the four components can be considered as a mixture, see \cite{raj95}.

In what follows, we also  define the transport velocities. Since we know that EPS encompasses the other cells, we can make the hypothesis that cyanobacteria, dead cells, and EPS have the same  solid velocity $\vs$, whereas the liquid has its own velocity    $\vl$. 
\subsection{Mass balance equations}

The equations for  the mass balance can be written as:
\begin{subequations}
\begin{align}
    \partial_{t} B +\nabla \cdot (  B \vs ) = \Gamma_B,
   \label{bm1} \\
    \partial_{t} D +\nabla \cdot (  D \vs ) = \Gamma_D, 
  \label{bm2}   \\
    \partial_{t} E +\nabla \cdot (  E \vs ) = \Gamma_E,
  \label{bm3}\\
 \partial_{t}  L+\nabla \cdot ( L \vl ) = \Gamma_L,
 \label{bm4}
\end{align}
\label{bm}
\end{subequations}
where  $\Gamma_{\phi}$, with $\phi = B,D,E,L$, are the reaction terms. 

Since we can find  only cyanobacteria, EPS and dead cells in the liquid, we work with  the following volume constraint: 
\begin{equation}
\label{vol1}
B+D+E+L=1.
\end{equation} 

Moreover,  from the mixture theory,  the total mass preservation of the mixture gives us the following equation~:
\begin{equation}
\label{eq1}
\Gamma_B + \Gamma_D + \Gamma_E + \Gamma_L = 0.
\end{equation}
Adding the equations of system \eqref{bm}, and using eq. \eqref{vol1} and \eqref{eq1}, we get the condition :
\begin{equation}
\label{condDiv}
\nabla \cdot \left( (1-L) \vs + L \vl \right) = 0,
\end{equation}
which can be seen as an averaged incompressibility.

\subsection{Force balance equations}

Now, to close  system \eqref{bm} and to find an equation for the velocities, we consider some force balance equations.
The  force balance equation for the component $\phi$ ($\phi=B,D,E,L$) is given by
\begin{equation}\label{force1}
\partial_{t} ( \phi \vu_{\phi}) +\nabla \cdot (   \phi \vu_{\phi} \otimes \vu_{\phi})=\nabla \cdot \tilde \T_{\phi}+\tilde \m_{\phi }+ \Gamma_{\phi} \vu_{\phi},
\end{equation}
where $\ds \tilde{\T}_{\phi}$ is the partial stress tensor and $\ds \tilde \m_{\phi }$ is the interaction with the other phases. 
The total conservation of momentum yields
\begin{equation}
\label{force1bis}
\sum_{\phi=B,D,E,L} (\tilde \m_{\phi }+ \Gamma_{\phi} \vu_{\phi})=0,
\end{equation}
which  means that the net momentum supply to the mixture due to all  components is equal to zero. As a matter of fact, if the mixture is closed, it is possible to prove that the sum of interaction forces and momentum transfers due to  mass exchanges is null.
Following \cite{cdnr}, we also  decompose the partial stress tensor as  $\ds \tilde{\T}_{\phi}=-\phi P I +\phi \T_{\phi}$, where  $\ds \T_{\phi}$ is the excess stress tensor and the interaction forces as $\ds \tilde \m_{\phi}= \m_{\phi}+ P \nabla \phi$.
Equation \eqref{force1} can therefore be rewritten as:
\begin{equation}
\label{force2}
\partial_{t} ( \phi \vu_{\phi}) +\nabla \cdot (   \phi \vu_{\phi} \otimes \vu_{\phi})= \m_{\phi}-\phi \nabla P+\nabla \cdot (\phi \T_{\phi})+ \Gamma_{\phi} \vu_{\phi}.
\end{equation}
Using equations \eqref{vol1}  
and \eqref{force1bis}, we find 
\[ \sum_{\phi\neq L} \m_\phi+ \Gamma_{\phi} \vu_{\phi}=-\m_L-\Gamma_{L}\vl\]
and therefore,  summing equations \eqref{force2} for $\phi=B,D,E$ and using \eqref{vol1} once again, we obtain 
\begin{equation*}
\partial_{t} ( (1-L) \vs) +\nabla \cdot (  (1-L) \vs \otimes \vs)= -(1-L) \nabla P+\nabla \cdot ( \sum_{\phi\neq L} \phi\T_{\phi}) -\m_L- \Gamma_{L} \vl
\end{equation*}
for the solid phase, whereas we have
\begin{equation*}
\partial_{t} ( L \vl) +\nabla \cdot (  L\vl \otimes \vl)= -L \nabla P+\nabla \cdot (\ds  L\T_{L}) +\m_L+ \Gamma_{L} \vl,
\end{equation*}
for the liquid phase. 

We  assume that the excess stress tensor is only present in the  equation for the solid  components $B$, $D$ and $E$, while in the equation for the liquid $L$,  only the hydrostatic pressure remains. 
This type of assumption is usually used in the theory of deformable porous media, where the excess stress tensor $\T_{L}$ is neglected in order to get Darcy like laws. 
More precisely, we  define the excess stress tensors by :
\begin{equation*}
\sum_{\phi\neq L} \phi\T_{\phi}=\Sigma \one \quad \textrm{and} \quad \T_{L}=0,
\end{equation*}
where $\Sigma$ is a monotone decreasing scalar function depending on the total solid volume ratio $B+D+E=1-L$. As a first approximation, useful for numerical tests, we take $\Sigma$ as  a linear  stress function defined by
\begin{equation} 
\label{sigma}
\Sigma=-\gamma (1-L), \textrm{ with } \gamma>0, 
\end{equation}
where negative values of $\Sigma$ indicate compression.

We also make the hypothesis  that the interaction forces for the liquid follow the Darcy law, that is to say we take $ \m_{L}$ proportional to the difference between the relative velocities
:
\begin{equation*}
 \m_{L}=-M(\vl-\vs), 
\end{equation*}
where $M$ is an experimental constant.

All these assumptions are made following works of  \cite{pre2},  \cite{astanin} and \cite{byrne}, where the theory of mixture is used to model tumour growth.
Thanks to these choices, we can rewrite the equations for the velocities as:
\begin{equation}\begin{array}{rl}\label{forcebis2}
\partial_{t} ( (1-L) \vs) +\nabla \cdot (  (1-L) \vs \otimes \vs)&= -(1-L) \nabla P+\nabla \Sigma  +M(\vl-\vs)- \Gamma_{L} \vl,\\ \\
\partial_{t} ( L \vl) +\nabla \cdot (  L\vl \otimes \vl)&= -L \nabla P-M(\vl-\vs)+ \Gamma_{L} \vl.
\end{array}\end{equation}

\subsection{Reaction terms}
Now the biological features of the system are included through the reaction terms, modeling the change of one phase to another, for example
 the growth and the death of involved organisms. We have to take into account the preservation of  the total mass, the life-cycle in this model being given by water-matter-water. The growth rates are denoted  $k_B$, $k_E$ and $k_D$, for cyanobacteria, EPS and dead cells respectively. We  consider that water is essential  for the growth of  cyanobacteria and EPS 
and for this reason their rate is multiplied by $L$; also,  EPS is produced by cyanobacteria so its growth rate is proportional to $B$. Finally, the dead cyanobacteria are transformed  in a part of dead cells $\alpha k_D$ and  a part of  water, due to their high water concentration.
The death rates  are denoted by  $k_D$, $\eps$ and $k_N$ for cyanobacteria, EPS and dead cells respectively. To sum up, the reaction terms can be written as : 
\begin{subequations}
\begin{align}
\Gamma_B&= k_B B L - k_D B,\\
 \Gamma_D&= \alpha  k_D B - k_N D, \\
  \Gamma_E&= k_E B L - \eps E.
\end{align}%
Eventually, we find the expression of  the mass exchange rate of liquid $\Gamma_L$ by condition \eqref{eq1}, that is to say :
\begin{equation}
\Gamma_L = B \left((1-\alpha)k_D - k_B L - k_E L\right) + k_N D +\eps E.
\end{equation}
\label{eq2}
\end{subequations}
Thus, equations \eqref{bm}, \eqref{condDiv}, \eqref{forcebis2} and \eqref{eq2} give the following closed system
\begin{equation}
\label{systemfinal}
\left\{\begin{array}{ll}
&\partial_{t} B +\nabla \cdot (  B\vs)= B\left(Lk_B(I,T,N)-k_D(I,T,N) \right),\\ \\
 &   \partial_{t} D +\nabla \cdot (  D\vs)=    \alpha B k_D(I,T,N)-D k_N(T), \\  \\
 &   \partial_{t} E +\nabla \cdot (  E\vs)= BL k_E (I,T,N) - \epsilon E,\\ \\
& \partial_{t}  L+\nabla \cdot ( L \vl)=B((1-\alpha)k_D(I,T,N)-L k_B(I,T,N)
 -L k_E (I,T,N))+D k_N(T)+\epsilon E, \\ \\
&\partial_{t} ( (1-L) \vs)  + \nabla \cdot (  (1-L) \vs \otimes \vs)+(1-L) \nabla P =  \nabla \Sigma
+(M-\Gamma_{L})\vl-M\vs,\\ \\
&\partial_{t} ( L \vl) +\nabla \cdot (  L\vl \otimes \vl)+L \nabla P=-(M-\Gamma_{L})\vl-M\vs,\\ \\
&\nabla \cdot \left( (1-L)\vs \right.  + L \left. \vl\right) =0.
\end{array}\right.
\end{equation}
We complement this system with  Neumann boundary conditions for the components
\begin{equation*}
\nabla B \cdot n|_{\partial \Omega}=\nabla E \cdot n|_{\partial \Omega}=\nabla D \cdot n|_{\partial \Omega}=0,
\end{equation*}
and no-flux boundary conditions for the velocities
\begin{equation}
\label{bound2}
\vs \cdot n|_{\partial \Omega}=\vl \cdot n|_{\partial \Omega}=0 .
\end{equation} 

\section{Stability of the constant stationary solutions in the 1D case}

In this section, we concentrate on the 1D case, ignoring the dependence on the light intensity and the temperature, and we consider therefore constant coefficients. We study  in this context the linear stability of the full system \eqref{systemfinal} around the non trivial equilibrium of the system. 

\subsection{Simplification of the system in 1D}

 In the particular case of dimension $1$ , we may simplify some equations using the boundary conditions. More precisely, we integrate  the incompressibility condition \eqref{condDiv}
complemented by the boundary conditions \eqref{bound2} and we obtain the following relation~:
\begin{equation*}
 (1-L)\vs  + L  \vl =0
 \end{equation*}
 or equivalently, using the fact that $L\neq 0$,
 \begin{equation}
\label{velo}
 \vl= \frac{L-1}{L}\vs,
 \end{equation}
Now, we add the two equations \eqref{forcebis2} on the velocities  in the 1D case and we use the last equation \eqref{velo}  to  obtain
 a more tractable expression for the derivative of $P$ as~:
\begin{equation}
\label{pre}
\partial_x P = - \gamma \partial_x (1-L) - \partial_x \left( \left(1-L\right) \vs^2 + L \vl^2 \right),
\end{equation}
where $\Sigma$ has been replaced by expression \eqref{sigma}. 
Substituting equation  \eqref{velo} in equation \eqref{pre} and expanding the space derivative, we obtain
\begin{equation}
\label{pre1}
\partial_x P = - \gamma \partial_x (1-L) - \partial_x \left( \frac{1-L}{L} \vs^2 \right) = - \left( \gamma + \frac{\vs^2}{L^2} \right) \pa_x(1-L) - 2 \frac{(1-L)}{L} \vs \pa_x \vs.
\end{equation}


%

Now, we add the sum of the first three equations of \eqref{systemfinal} on $B,D,E$  and we multiply it by $\vs$ to obtain~:
\begin{equation*}
\vs\pa_{t}  (1-L)+ \vs\pa_{x}((1-L)\vs)=-\Gamma_{L} \vs.
\end{equation*}
Using this last equation  in the expansion of   the first equation of \eqref{forcebis2} on $\vs$
and dividing by $(1-L)$,  in the case when $L \ne 1$, we obtain
\begin{equation*}
\pa_t \vs + \vs \pa_x \vs + \partial_x P + \frac{\gamma}{1-L} \pa_x (1-L) = 
\frac{(\Gamma_L - M)}{1-L} (\vs - \vl).
\end{equation*}
Now, we  simplify this last equation in the case when $L \ne 0$, using equations \eqref{pre1} and \eqref{velo} as
\begin{equation*}
\pa_t \vs + \frac{3L-2}{L} \vs \pa_x \vs + \left( \frac{L}{1-L} \gamma - \frac{\vs^2}{L^2} \right) \pa_x (1-L) = \frac{\Gamma_L - M}{L(1-L)}\vs.
\end{equation*}
Finally,  when $L \ne 0, 1$,  system \eqref{systemfinal}  in the 1D case  reduces to  the following equations
\begin{subequations}
\begin{align}
&   \partial_{t} B + \vs \pa_x B + B \pa_x \vs = k_B B L - k_D B,
 \\
  & \partial_{t} E + \vs \pa_x E + E \pa_x \vs = k_E B L - \eps E, 
  \\
   &\partial_{t} D + \vs \pa_x D + D \pa_x \vs = \alpha  k_D B - k_N D,
 \\
 & L=1-(B+D+E), \\
 & \pa_t \vs + \frac{3L-2}{L} \vs \pa_x \vs + \left( \frac{L}{1-L} \gamma - \frac{\vs^2}{L^2} \right) \pa_x (B+E+D) = \frac{\Gamma_L - M}{L(1-L)}\vs,
 \\
 & \vl= \frac{L-1}{L}\vs. 
\end{align} 
\label{system1D}
\end{subequations}
with the boundary conditions 
\begin{equation}\label{bdy1D}
\pa_x B = \pa_x E = \pa_x D = \vs = 0 \textrm{ in }x=0 \textrm{ and in }x=1. 
\end{equation}

\subsection{Stationary states}
All the stationary states should satisfy the following relation and inequalities : $B+D+E+L = 1$ and $0 \leq B,E,D,L \leq 1$.

We first notice from equations \eqref{systemfinal} that 
\begin{equation}\label{stat1}
B=D=E=0, \quad L=1, \quad \vs=\vl=0 
\end{equation}
is a straightforward constant stationary solution. Moreover, $L=0$ would not lead to a constant stationary solution and
 we may assume in what follows that $L\neq 0$.
 
  The other homogeneous steady state  of the system  \eqref{systemfinal}, is given by 
\begin{equation}
\label{stat2}
 \bar{B} = \ds\frac{\ds 1 - \frac{k_D}{k_B}}{\ds 1 + \alpha \frac{k_D}{k_N} + \frac{k_E \cdot k_D}{\varepsilon \cdot k_B}}, \quad
  \bar{E} =\ds \frac{k_E  k_D}{\varepsilon k_B} \bar{B}, \quad
   \bar{D} =\ds \alpha \frac{k_D}{k_N} \bar{B}, \quad
   \bar{L} = \frac{k_D}{k_B}, \quad  \vs=\vl= 0.
\end{equation}
 
 \subsection{Linear stability of the stationary states for the ODE system}

To begin with, we study the linear stability of the stationary states \eqref{stat1} and \eqref{stat2} with respect to  the source  part of  system \eqref{system1D}. To this aim, we  use the classical  Routh-Hurwitz conditions for stability.

Assuming that all coefficients are constant, we consider the following system
\begin{subequations}
\begin{align}
    \partial_{t} B &= k_B B L - k_D B,
   \label{lin1} \\
    \partial_{t} E &= k_E B L - \eps E, 
  \label{lin2}   \\
    \partial_{t} D &= \alpha  k_D B - k_N D,
  \label{lin3}
\end{align}
\label{lin}
\end{subequations}
with $k_{B}, k_{D}, k_{E}, \eps >0$.
We denote  $\bar{W} = (\bar{B},\bar{E},\bar{D})$ and we linearize system  \eqref{lin}  in $\bar{W}$, which yields
\begin{equation*}
W_t = J(\bar{W}) (W - \bar{W}),
\end{equation*}
where $J(\bar{W})$ is the jacobian computed at $\bar W$.
The characteristic polynomial of $J(\bar{W})$ can be written as
\begin{equation*}
P(\lambda) = a_{3}\lambda^3 + a_2 \lambda^2 + a_1 \lambda + a_0, 
\end{equation*}
and the  Routh-Hurwitz necessary and sufficient conditions, see \cite{mur1},  to prove the stability of  the stationary solution 
$\bar{W} $
 are given by~:
\begin{equation}
\label{rh}
  a_n > 0,  \, \forall n \in \{0, \cdots, 3 \} \quad
  a_1 a_2 - a_{0}a_3 > 0.
\end{equation} 
  
\begin{proposition} 
\begin{itemize}
\item The stationary solution \eqref{stat1} is stable for system \eqref{lin}  iff $k_{D}> k_{B}$.
\item The stationary solution \eqref{stat2}  is always stable for system \eqref{lin}.
\end{itemize}
\end{proposition}
The first statement means that water becomes everywhere dominant and no biofilm is formed if and only if the  death rate of the cyanobacteria is larger than their birth rate. 

\begin{proof}

We consider first the stationary solution \eqref{stat1}. The jacobian matrix is equal to 
\begin{equation*}
J(\bar{W}) = 
\left( \begin{array}{c c c}
 k_B-k_{D} &  0 & 0 \\
 k_E   & - \varepsilon & 0 \\
 \alpha k_D  & 0   & - k_N
\end{array} \right)
\end{equation*}
It is obvious here that the three eigenvalues are $- \varepsilon <0$, $ - k_N<0$ and $ k_B-k_{D}$. Therefore, the stationary solution \eqref{stat1} is stable iff  $ k_B-k_{D}<0$.

Now, in the case of the stationary solution \eqref{stat2}, the jacobian matrix is equal to 
\begin{equation*}
J(\bar{W}) = 
\left( \begin{array}{c c c}
 - k_B \bar{B} & - k_B \bar{B} & - k_B \bar{B} \\
 k_E (\bar{L} - \bar{B})   & - k_E \bar{B} - \varepsilon & - k_E \bar{B} \\
 \alpha k_D  & 0   & - k_N
\end{array} \right)
\end{equation*}
and we can compute the coefficients of the polynomial : 
\begin{equation*}
\left\{ \begin{array}{l}
a_{3}=1, \\
  a_2 = \bar{B}( k_B + k_E) +( k_N + \varepsilon), \\
  a_1 = \bar{B} (k_B k_N +  k_E k_N +  k_D k_E +   k_B k_D \alpha+  k_B \varepsilon )+ k_N \varepsilon, \\
  a_0=  \bar{B}( k_D k_E k_N +  k_B k_N \varepsilon + k_B k_D \alpha \varepsilon). \\
\end{array} \right.
\end{equation*}
The first two conditions of the Routh-Hurwitz conditions \eqref{rh} are obvious and  the third condition 
is satisfied,  observing that the three terms of $a_{0}a_3$ are contained in $a_1 a_2$ and that  $a_1 a_2$   is a sum of  positive terms.
\end{proof}

 \subsection{Linear stability of the stationary states for the hyperbolic system \eqref{system1D}}

Now, we want to study the stability of the stationary solution  for the  full hyperbolic system.
We can rewrite  system  \eqref{system1D}  under the vectorial form 
\begin{equation}
\label{systemVect}
U_t + A(U) U_x = \Gamma_U,
\end{equation}
 where
\begin{equation*}
U = 
\left( \begin{array}{l}
 B \\
 E \\
 D \\
 \vs \\
\end{array} \right),\,\,
A(U) = 
\left( \begin{array}{c c c c}
 \vs & 0   & 0   & B \\
 0   & \vs & 0   & E \\
 0   & 0   & \vs & D \\
 \eta & \eta & \eta &\ds  \frac{3L-2}{L}\vs \\
\end{array} \right),\,\,
\Gamma_U = 
\left( \begin{array}{c}
 k_B B L - k_D B \\
 k_E B L - \eps E \\
 \alpha  k_D B - k_N D \\
\ds \frac{\Gamma_L - M}{L(1-L)}\vs \\
\end{array} \right) 
\end{equation*}
with $\ds \eta = \left( \frac{L}{1-L} \gamma - \frac{\vs^2}{L^2} \right)$ and $L=1-(B+E+D)$.  


Linearizing  system \eqref{systemVect} around a  homogeneous steady state $\bar U$ with $\vs=0$,  we obtain
\begin{equation}
\label{linf}
\pa_t w + A(\bar{U}) \pa_x w = \Gamma_U'(\bar{U}) w,
\end{equation}
where
\begin{equation*}
A(\bar{U}) = 
\left( \begin{array}{c c c c}
 0   & 0   & 0   & \bar B \\
 0   & 0   & 0   & \bar E \\
 0   & 0   & 0   & \bar D \\
 \ds \frac{\bar L}{1-\bar L} \gamma & \ds \frac{\bar L}{1-\bar L} \gamma  &  \ds \frac{\bar L}{1-\bar L} \gamma  & 0 \\
\end{array} \right)
\end{equation*}
and
\begin{equation*}
\Gamma_U'(\bar{U}) = 
\left( \begin{array}{cccc}
 k_B  (\bar{L} - \bar{B}) -k_{D}& - k_B \bar{B} & - k_B \bar{B}  & 0 \\
 k_E (\bar{L} - \bar{B})   & - k_E \bar{B} - \varepsilon & - k_E \bar{B} & 0 \\
 \alpha k_D  & 0   & - k_N &  0\\
0  &  0  &  0  & \ds \frac{-M}{\bar{L}(1-\bar{L})} \\
\end{array} \right),
\end{equation*}
using the fact that for the stationary solutions we consider $\vs=0$ and $\Gamma_{L}=0$ .

Owing to boundary conditions \eqref{bdy1D}, we are considering solutions  of Fourier type under the following form :
\begin{equation}
\label{sol1}
\left\{ \begin{array}{l}
w_b =\ds \sum_{n} B_0^n e^{\lambda_n t} \cos(\pi n x),\\
w_e = \ds\sum_{n} E_0^n e^{\lambda_n t} \cos(\pi n x),\\
w_d = \ds \sum_{n} D_0^n e^{\lambda_n t} \cos(\pi n x),\\
w_{\vs} =\ds \sum_{n} v_0^n e^{\lambda_n t} \sin(\pi n x),
\end{array} \right.
\end{equation}
where $0 \leq x \leq 1$.
Inserting  solutions \eqref{sol1} in equation \eqref{linf} and using the expression  $\ds\bar{L} = \frac{k_D}{k_B}$ , we are reduced to find the sign of the eigenvalues of matrix
\begin{equation}
\label{matrix}
A = \left( \begin{array}{cccc}
 -k_B \bar{B} & -k_B \bar{B} & -k_B \bar{B} & -N \bar{B} \\
 k_E (\bar{L} -\bar{B}) & -k_E \bar{B} - \varepsilon & -k_E \bar{B} & -N \bar{E} \\
    \alpha k_D & 0 & -k_N & -N \bar{D} \\
  \ds  \bar \eta  N & \ds   \bar \eta  N & \ds  \bar \eta  N &\ds -\frac{M}{\bar{L}(1-\bar{L})} \\
\end{array} \right),
\end{equation}
where $N=\pi n$ and $\ds \bar \eta = \frac{\bar L}{1-\bar L} \gamma $. 
Let us define the characteristic polynomial
\begin{equation*}
\label{cpol}
P(\lambda) = a_4 \lambda^4 + a_3 \lambda^3 + a_2 \lambda^2 + a_1 \lambda + a_0.
\end{equation*}
For a fourth-order polynomial under the previous form, the Routh-Hurwitz conditions read as~:
\begin{equation}\label{rh2}
    a_n > 0,  \, \forall n \in \{0, \cdots, 4 \} \quad
    a_3a_2 > a_4a_1, \quad
    a_3a_2a_1 > a_4a_1^2 + a_3^2a_0.
\end{equation}
Using some relations \eqref{stat2}, the coefficients related to matrix \eqref{matrix} may be written as :
\begin{align*}
a_0& = \frac{ M }{\bar{L}(1 - \bar{L}) }\bar{B} \left( k_E k_N  k_D+ k_N k_B\varepsilon + k_D  k_B \alpha \varepsilon \right) +
 N^2 \bar \eta \bar{B} \left(\frac{k_E k_N k_D}{k_B} + k_N \varepsilon + k_D \alpha \varepsilon \right), \\
a_1 & =  
\frac{ M }{ \bar{L}(1 - \bar{L})}\bar{B}  \left( k_B k_N + k_E k_N + k_E k_D + k_B k_D \alpha + k_B \varepsilon \right)+\frac{ M }{ \bar{L} (1- \bar{L})}  k_N \varepsilon
 \\& \quad \quad \quad \quad \quad
+N^2 \bar \eta \bar{B} (k_{N}+\varepsilon)\left( 1+\alpha\frac{k_{D}}{k_{N}}+\frac{k_{E}k_{D}}{\varepsilon k_{B}}\right)
+ \bar{B}  \left( k_E k_N k_{D} + k_B k_N \varepsilon + k_B k_D \alpha \varepsilon \right) , \\
%
a_2 &=  \frac{ M }{\bar{L} (1- \bar{L})} \bar{B}( k_B +  k_E )+ 
 \frac{ M }{\bar{L} (1- \bar{L})}\left(k_N + \varepsilon \right) +  N^2 \bar \eta  \bar B \left( 1 + \ds \frac{k_E  k_D}{  k_B\varepsilon} + \frac{k_D}{k_N} \alpha \right)+\bar{B} ( k_B k_N + k_E k_N \\
& \qquad \qquad   + k_E k_D +  k_B k_D \alpha+ k_B \varepsilon ) + k_N \varepsilon,
 \\ 
a_3 &= \frac{M}{\bar{L} (1- \bar{L})} +\bar{B}( k_B + k_E)  +  (k_N+ \varepsilon), \\
a_4 & = 1.
\end{align*}
Now, let us control the Routh-Hurwitz conditions \eqref{rh2}. The first condition $a_{n}>0$ for all $n$  is obvious, since they
are all  composed of positive terms.

The second term is show to be positive by noticing that the  $15$ terms of the product  $a_1a_4=a_{1}$ are all contained in the $63$ terms of the product $  a_2a_3$. The $48$ remaining terms, all positive,  may be written as :
\begin{equation}\label{expr}
\begin{split}
a_{2} a_{3}-a_{1}a_{4}&= \left(\frac{ M }{\bar{L} (1- \bar{L})}\right)^2 \left( \bar B (k_{B}+k_{E})+(k_{N}+\varepsilon)\right)+ 
\frac{ M }{\bar{L} (1- \bar{L})}  N^2 \bar \eta \bar B \left( 1 + \ds \frac{k_E  k_D}{  k_B\varepsilon} + \frac{k_D}{k_N} \alpha \right)\\
&+ \frac{ M }{\bar{L} (1- \bar{L})}\left( \bar B (k_{B}+k_{E})+(k_{N}+\varepsilon)\right)^2+N^2 \bar \eta \bar B^2(k_{B}+k_{E})
\left( 1 + \ds \frac{k_E  k_D}{  k_B\varepsilon} + \frac{k_D}{k_N} \alpha \right) \\
&+ \bar B^2 \left( (k_{B}+k_{E})(k_{B}k_{N}+k_{E}k_{D}+k_{E}k_{N}+ k_{B}k_{D}\alpha+k_{B}\varepsilon )\right)  \\
&+ \bar B \left( (k_{B}+k_{E})(k_{N}^2+ 2k_{N}\varepsilon)+ k_{B}k_{D}k_{N}\alpha+ k_{E}k_{D}\varepsilon+ k_{B}\varepsilon^2\right)
+ k_{N}\varepsilon(k_{N}+\varepsilon).
\end{split}
  \end{equation}
  Now, let us notice that the last condition can be rewritten as $\ds  (a_3a_2 - a_4a_1)a_1 > a_3^2a_0$. Using this last expression \eqref{expr}, it is enough to show that the $150$ terms of  $a_3^2a_0$ are all included in the $720$  terms of the product $ (a_3a_2 - a_4a_1)a_1$. This can be done by checking, as we did, each of the following $12$ different  categories of  terms:  
 the terms with $\ds \bar B \left(\frac{ M }{\bar{L} (1- \bar{L})}\right)^3$ in factor, with $\ds \bar B^2 \left(\frac{ M }{\bar{L} (1- \bar{L})}\right)^2$, with $\ds \bar B \left(\frac{ M }{\bar{L} (1- \bar{L})}\right)^2$, 
 with $\ds\bar B^3 \frac{ M }{\bar{L} (1- \bar{L})}$, with $\ds \bar B^2 \frac{ M }{\bar{L} (1- \bar{L})} $, with $\ds \bar B \frac{ M }{\bar{L} (1- \bar{L})}$, 
  with $\ds  \bar B\left( \frac{ M }{\bar{L} (1- \bar{L})}\right)^2 N^2 \bar \eta$, with $\ds  \bar B^2 \frac{ M }{\bar{L} (1- \bar{L})} N^2 \bar \eta$, with $\ds \bar B \frac{ M }{\bar{L} (1- \bar{L})} N^2 \bar \eta$, with $\ds  \bar B^3  N^2 \bar \eta$, with $\ds  \bar B^2 N^2 \bar \eta$, with $\ds \bar B N^2 \bar \eta$.

This result can be summarized in the following proposition~:
{\begin{proposition}
 The stationary solution \eqref{stat2}  is always stable for the full hyperbolic system \eqref{systemVect}
\end{proposition}}

\section{Influence of the environment  on the biofilm growth}

In the previous section, we were dealing with the case of constant coefficients and we study now the addition of a dependence on light and temperature  in our model.
All the coefficients of the system may depend on temperature, light intensity and concentration of nutrients. 
However, the main factor which influences the life of photoautotrophic cyanobacteria is the light, allowing these organisms to photosynthesize inorganic compounds. Environmental light intensity variation potentially accounts for much of the variation in the physiology and population growth of cyanobacteria. 
Moreover, there exists a range of temperatures, as well as a range of nutrient concentrations, necessary to the survival of  the cyanobacteria.
Here, we consider that the cyanobacteria have sufficient supplies of nutrients, neglecting thus  its influence. 

To estimate the growth rate of cyanobacteria, we decompose  the coefficient $k_B$ as follows
\begin{equation}
\label{geq1}
k_B = k_{B0}\; g(I,T),
\end{equation}
where $k_{B0}$ is the optimal growth rate of cyanobacteria and $g(T, I) \in [0,1]$ is an efficiency factor which takes into account the influence of temperature and light. Many authors have already formulated the effect of light upon algae growth with empirical mathematical functions, see \cite{theb03}. More precisely, in many models, effects of light and temperature are assumed  to be independent, and the resultant efficiency factor  is thus  taken as the product of two limiting factors, namely $g_T(T)$  for the temperature and $g_I(I)$ for the light. Therefore, in absence of appropriate experiments, 
 we assume here that
\begin{equation}
\label{geq2}
g(I,T) = g_I (I) \cdot g_T (T).
\end{equation}

\subsection{Light dependence}

Let us begin with modeling  the light influence on the growth rate. 
Since phototropic organisms use specific parts of the light spectrum, the light absorption in the  biofilm  depends on the specific species of organisms  we study in the biofilm. For example, it has been shown that different kinds of cyanobacteria, such as red and green cyanobacteria,  can coexist by absorbing different parts of the light spectrum, see \cite{sto07a}. As a  first step,  we have to estimate the light absorption by clear water. The absorption of water's overtone bands within the visible spectrum are quite small, varying between $0.3-0.01$ $(m^{-1})$, see the interesting work of \cite{sto07} and  the absorption coefficient depends on the wavelength of the incident light. 
Let us s denote by $I_0(t)$ the incident light intensity on the upper surface of water, and by $I(x,y,t)$ the light intensity in the water. Also,  the light intensity is attenuated following the law of photon absorption in the matter, i.e. on the vertical axes. We call $y \in [0,H]$  the vertical length    and we have:
\begin{equation*}
\frac{I(y,t)}{I_0(t)} = e^{-\int_0^y \mu (s) ds},
\end{equation*}
where the absorption coefficient $\mu$ depends on the matter and on the frequency of radiation. 

Now, we assume that the absorption coefficient $\mu$ is linear in the volume fractions, that is to say it has the following form : 
\begin{equation*}
\mu = \mu_{0} \left( 1 + \mu_{B} \left( B + E + D \right) \right),
\end{equation*}
where $\mu_0$ is the absorption coefficient when the water is clear and $\mu_{B}$ is a second absorption coefficient in presence of biomasses. To choose the value of $\mu_0$, we consider some experimental observations on biofilm growth, which are realised using "Truelight lamps" (Auralight, Sweden), see \cite{zip07}, \cite{alb08} and \cite{dip09}. These kinds of lamps have two maxima of irradiation, the first one  at a wavelength of about $550$ $(nm)$ and the second one at  a wavelength of about $620$ $(nm)$, see \cite{zip07}. Now, following   \cite{sto07}, we consider  that the light absorption by water at the wavelength near the first maximum is nearly equal to  $ 0.09 $ $(m^{-1})$, while the absorption corresponding to the second maximum, which is near the fifth harmonic of the symmetric and asymmetric stretch vibrations, has a value of $0.2$ $(m^{-1})$. Thus, we choose an intermediate value between them: 
\begin{equation}
\label{mu0}
\mu_0 \approx 0.1 \,\,\,(m^{-1}).
\end{equation}
To estimate the value of $\mu_B$, we follow some experimental observations in \cite{zip07}. In this work, the initial (without biofilm) transmitted light is about $95 \%$; at the mature stage, after around $30$ days, the light transmitted throughout biofilm is about $10 \%$ of the incident light. At the same time, the biofilm thickness measured in several cases varies between $0.4-0.6$ $(mm)$. From  these observations, we estimate  that for  a  biofilm thickness equal to  
$h=0.06$ $(cm)$, 
we have a ratio $I/I_0 = 15 \%$
, that is to say
\begin{equation*}
\frac{I}{I_0} = e^{- \mu_0 \mu_B h} = 0.15.
\end{equation*}
Taking $\mu_0 = 0.001$ $(cm^{-1})$, see eq. \eqref{mu0}, and $h = 0.06$, we obtain the following estimate  $\mu_B = -log(0.15)/(\mu_0 h) \approx 3 \cdot 10^4$. 

\begin{figure}
\centering
  \includegraphics[width=10cm, height=5cm]{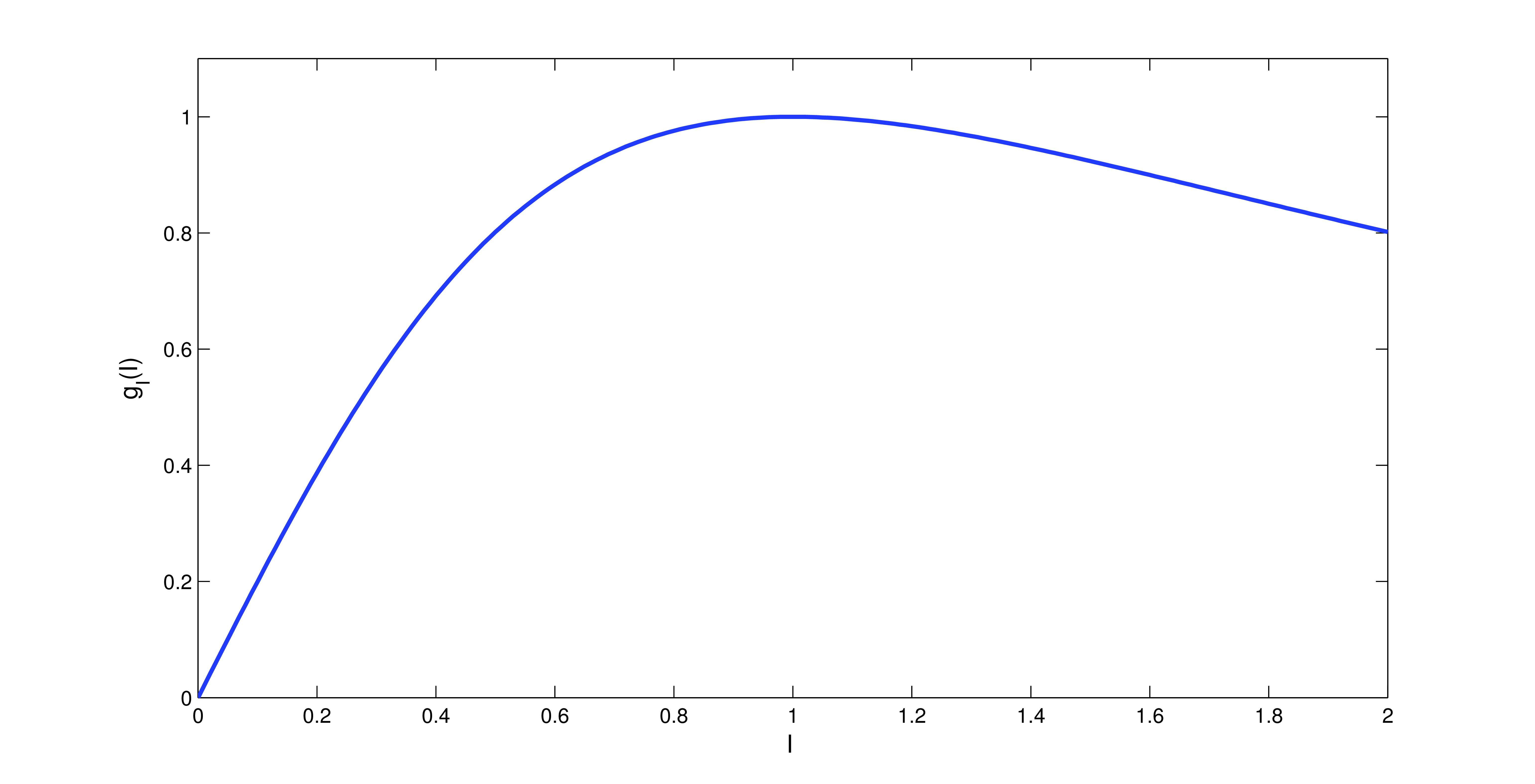}
\vskip-0.1 cm
\caption{Function $g_I(\hat{I})$.}
\label{gI} 
\end{figure}

Now, we express the limiting factor $g_{I}(I)$ of the growth rate $k_B$ as a function of the previous estimated absorption.
Following \cite{theb03}, \cite{eil88} and references therein, we assume that the specific growth rate as function of irradiation $I(x,y,t)$ is given by
\begin{equation*}
g_I (I)= 2 w_I \left(1 + \beta_I \right) \frac{\hat{I}}{\hat{I}^2 + 2\beta_I \hat{I} + 1},  
\end{equation*}
where $\hat{I} = I/I_{opt}$, and $I_{opt}$ is the optimal light for the  cyanobacteria evolution.  In this formula,  $w_I$ is the maximum specific growth rate and $\beta_I$ is a shape coefficient. The maximum is reached for $I = I_{opt}$ which implies $w_I = 1$, and considering the works of  \cite{dip09} and \cite{alb08}, we estimate that the optimal growth  is obtained in our case with an incident light approximately equal to   $I_{opt} = 0.01$ $(\mu mol\, cm^{-2}\, sec^{-1})$. Therefore, when $I_0 = 0.0015$ $(\mu mol\, cm^{-2}\, sec^{-1})$ the growth is very small;  for larger values  of $I_{0}$, the growth seems to be linear with light intensity;  when $I_0$ is greater than $0.01$ $(\mu mol\, cm^{-2}\, sec^{-1})$ there is a saturation effect and the growth rate diminishes.
 Studying the dependence  of  the growth  rate  with respect to light  as done in \cite{zip07}, we estimate that $\beta_I = 0.01$;  using the obtained values, we display  the function $I \to  g_{I}(I)$ in Figure \ref{gI}, with a maximum corresponding to $\hat{I} = I/I_{opt} = 1$.

\subsection{Temperature dependence}

Secondly, we want to quantify  the influence of temperature on the biofilm growth rate. For simplicity, we assume that all the  liquid is at the same temperature, neglecting the influence of the thermal capacity. 
We denote by  $T_{opt}$ the  optimal temperature for which  cyanobacteria have a maximal growth rate. As in the case of light, we make the hypothesis  that the growth rate diminishes when the temperature is far from its optimal value. Thus, following \cite{theb03}, we choose the specific growth rate $g_{T}(T)$ as a function of temperature $T$ in the following way~:
\begin{equation*}
g_T(T) = 2 w_T \left( 1 + \beta_T \right) \frac{\theta}{\theta^2 + 2 \beta_T \theta + 1},
\end{equation*}
where
\begin{equation}
\label{teq2}
\theta = \frac{T - T_{min}}{T_{opt} - T_{min}}.
\end{equation}
Here, $w_T$ corresponds to the maximal growth rate and considering eq. \eqref{teq2}, the optimal condition corresponds to $w_T = 1$. As in the previous case, $\beta_T$ is a shape parameter, and $T_{min}$ is the minimal temperature for the model. Reasonably, we assume $T_{min}=0$, and following \cite{dip09} and \cite{alb08} an optimal temperature could be  $T_{opt} = 30^o C$ with a coefficient  $\beta_T = 0.1 $.
Note that the function $g_T(\theta)$ has the same shape as $g_I(\hat{I})$.

\subsection{Parameters estimate}

The cyanobacteria growth rate in optimal condition ($k_{B0}$) can take  several values depending on the cyanobacteria species under observation. Since their doubling time seems to vary  between some hours up to some days, see for example \cite{joh98} and references therein, we assume, for simplicity, a doubling time of $1$ day, which means $k_{B0} = 8 \cdot 10^{-6}$ $(sec^{-1})$. 
Following the work \cite{dip09}, we can observe that the growth of biofilm after about $30$ days, that is to say up to the mature stage, is characterized by  a  thickness of an  order of magnitude of $1 \, mm$. A summary  of all parameters estimates  of model \eqref{systemfinal} is displayed in  Table \ref{partab}

\begin{table}[ht!]
\centering
       \begin{tabular}{|l|l|l|l|}
\hline
Parameter & Value & Unit of measurement & Indications  \\
\hline
$k_{B0}$ & $8 \cdot 10^{-6}$ & 1/$sec$ & Cyanobacteria growth rate  \\
$k_{E}$ & $12 \cdot 10^{-6}$ & 1/$sec$ & EPS growth rate \\
$k_{D}$ & $0.2 \cdot 10^{-6}$ & 1/$sec$ & Cyanobacteria death rate \\
$k_{N}$ & $1 \cdot 10^{-8}$ & 1/$sec$ & Dead cells consumption rate \\
$\epsilon$ & $1 \cdot 10^{-7}$ & 1/$sec$ & EPS death rate \\
$\alpha$ & $0.25$ & dimensionless & Fraction dead cells \\
$M$ & $1 \cdot 10^{-8}$ & 1/$sec$ & Tensor coefficient \\
$\gamma$ & $5 \cdot 10^{-16}$ & $cm^2/sec^2$ & Tensor coefficient \\
$\mu_0$ & $0.001$ & 1/$cm$ & Light absorption by water \\
$\mu_{B}$ & $3\cdot 10^{4}$ & dimensionless & Light absorption by biomasses \\
\hline
\end{tabular}
\vskip -0.1cm
\caption{List of (dimensional) parameters }\label{partab}
\end{table}


\section{Sensitivity and Robustness of parameters}
Our model includes many parameters and once we have determined a reference set of values, we have to study
their influence on the model dynamics by a sensitivity and robustness analysis, as explained in this section.
However, interpreting these results deserves some caution. Rather than meaning that a parameter is pivotal in the system, it could reflect a constructed parameter, which is hiding several more real parameters. In any case, sensitivity and robustness are a marker of something that should be explored more deeply.

\subsection{Sensitivity analysis}
In this subsection, we study numerically the sensitivity of the model to the parameters variations in the two dimensional case. It is important to check this point before using the model with experimental results, in order to be aware of the limitations due to parameter values used in the simulations. The sensitivity study shows how the behavior of the global system depends on each of its components. It can also give some information for the further exploration of the parameter space.

To this goal, we consider four outputs of the model: the total volume of cyanobacteria, EPS, dead cells and the whole biofilm; the parameters chosen are the temperature $T$, the light intensity on the upper surface of the water $I_0$, the tensor coefficients $\gamma$ and $M$, and the growth and death rates of the different components $k_{B0}$, $k_{E}$, $k_{D}$, $\epsilon$, $k_{N}$ and $\alpha$.
 
We proceed as follows: if $s(p_1, p_2,..., p_n)$ is one of the  outputs obtained with the parameter values, the sensitivity of this output for example to the parameter $p_1$ is given by:
\begin{equation*}
S = \left(\frac{s(p_1 \pm \delta, p_2,..., p_n) - s(p_1, p_2,..., p_n)}{s(p_1, p_2,..., p_n)}\right)/\left( \frac{\delta}{p_1} \right);
\end{equation*}
here, we consider a parameter change equal to $ \delta = 0.05 \cdot p_1$, i.e. $5 \%$ of the parameter $p_1$.
With reference to the parameters values, we use the ones reported in Table \ref{partab}. 


 Our domain is  the square $\Omega = [0,L]\times [0,L]$ where $L=5\, cm$.  We consider as an initial datum for the cyanobacteria in the numerical domain $\Omega^* = [0,1]\times [0,1]$ the following function   
\begin{equation}\label{initialdata2d}
B_0(x,y)=0.1\exp \left[ -\frac{\left(x-0.5\right)^2}{0.0005} \right]  \cdot \exp \left[ -\frac{y^2}{0.00004} \right].
\end{equation} The initial volume of cyanobacteria is equal to $ V_{B0} = 4.0711\cdot10^{-5} \, (cm^2) $, while the other components are initially zero.
We perform $30$ days simulations and we report the values obtained in Table \ref{tab1}. 
 \begin{figure}[!ht]
    \centering
        \includegraphics[height=10cm]{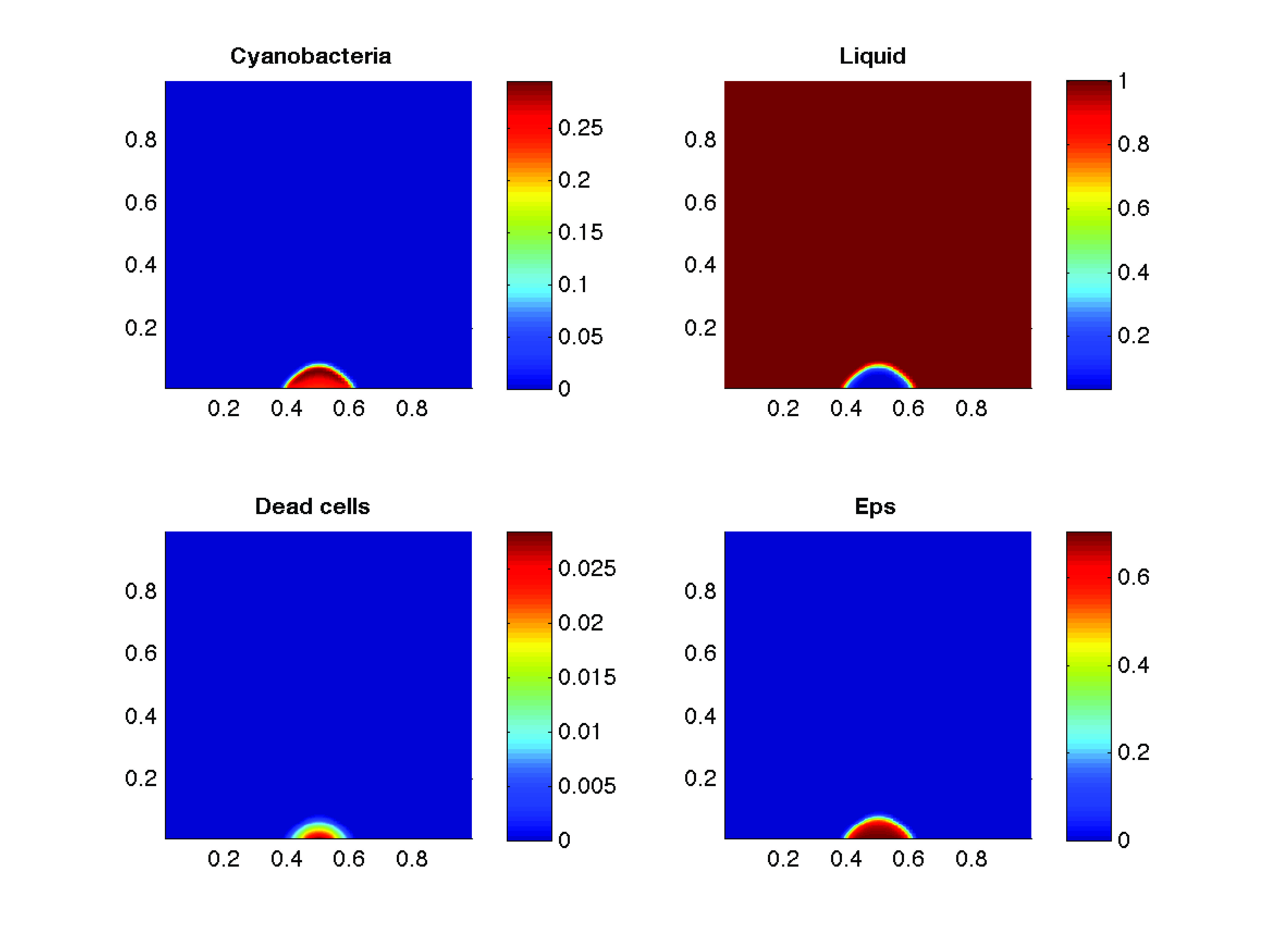}
    \vskip -0.1cm
    \caption{\footnotesize{Volume fractions of the biofilm components (cyanobacteria, dead cyanobacteria, EPS) and the liquid at $30$ days. Here the initial datum is \eqref{initialdata2d} for cyanobacteria while the other components are initially zero. With reference to the parameters, we consider the values reported in Table \ref{partab}.}}
    \label{bio30}
\end{figure}

\begin{table}[ht!]
\centering
     
\begin{tabular}{|l|l|l|l|l|l|l|l|l|}
\hline
Par. changed & $V_{bio}$ & $S_{bio}$ & $V_{B}$ & $S_{B}$ & $V_E$ & $S_E$ & $V_D$ & $S_D$  \\
\hline
$None$ & $ 0.2441$ & 0 & $0.0745$ & 0 & $0.1665$ & 0 & $0.0031$ & 0 \\
 \hline
$k_{B0}$ + $\delta$ &  $ +  8.26 \% $ & $     1.65  $ &  $ +11.39\% $ & $ 2.28 $ & $  + 6.76 \% $ & $   1.35$ & $ +13.54 \% $ & $   2.7$ \\ 

$k_{B0}$ - $\delta$ &  $   - 8.05 \% $ & $    1.61  $ &  $-10.84 \% $ & $  2.17 $ & $   -6.76 \% $ & $   1.35$ & $ - 10.28 \% $ & $    2.06 $ \\ 

$k_{E}$ + $\delta$ &  $+ 0.36 \% $ & $ 0.07$ &  $ -2.84\% $ & $  0.57 $ & $  + 1.83 \% $ & $  0.37$ & $ - 1.39 \% $ & $  0.28$ \\

$k_{E}$ - $\delta$ &  $ -0.35\% $ & $  0.07 $ &  $+2.99 \% $ & $ 0.60 $ & $  -1.93 \% $ & $  0.39 $ & $ +4.15  \% $ & $     0.83 $\\ 

$k_{N}$ + $\delta$ &  $   +0.013\% $ & $0.0027   $ &  $-0.013\% $ & $  0.0025 $ & $+ 0.002 \% $ & $ 0.0004$ & $ +1.26  \% $ & $  0.25 $ \\ 
$k_{N}$ - $\delta$ &  $  +0.014\% $ & $ 0.0028  $ &  $-0.013 \% $ & $   0.0026 $ & $  +0.0013 \% $ & $ 2.6 \cdot 10^{-4}$ & $ + 1.33\%$&$ 0.26 $ \\ 

$k_{D}$ + $\delta$ &  $ -0.27\% $ & $ 0.054  $ &  $ -0.92\% $ & $  0.184 $ & $  -0.094 \% $ & $     0.019$ & $ +5.57 \% $ & $ 1.115$ \\ 
$k_{D}$ - $\delta$ &  $ + 0.3\% $ & $  0.06  $ &  $   +0.9\% $ & $  0.181 $ & $  +   0.094 \% $ & $   0.019$ & $ -3.05  \% $ & $ 0.6104 $ \\ 
$\eps$ + $\delta$ &  $-0.053 \% $ & $ 0.011   $ &  $ +0.07\% $ & $ 0.014 $ & $    -0.13\% $ & $     0.027$ & $ +1.37  \% $ & $   0.274$ \\
$\eps$ - $\delta$ &  $ +0.08\% $ & $    0.016 $ &  $-0.095\% $ & $  0.019$ & $  +   0.14 \% $ & $    0.028$ & $ +1.22  \% $ & $    0.244 $ \\ 

$M$ + $\delta$ &  $  -0.13\% $ & $ 0.026 $ &  $-0.13 \% $ & $ 0.026 $ & $   -0.15 \% $ & $  0.03$ & $  +1.18  \% $ & $ 0.237$ \\ 
$M$ - $\delta$ &  $+0.16 \% $ & $    0.032 $ &  $+ 0.11 \% $ & $  0.028 $ & $ + 0.16 \% $ & $ 0.032$ & $ + 1.41  \% $ & $ 0.2821 $ \\ 

$\gamma$ + $\delta$ &  $+1.24\% $ & $0.247   $ &  $   + 0.97\% $ & $     0.195$ & $ +1.33 \% $ & $  0.267$ & $ +2.24  \% $ & $  0.449 $ \\ 
$\gamma$ - $\delta$ &  $ -1.24\% $ & $   0.248  $ &  $    -1.03\% $ & $  0.206 $ & $   -1.37 \% $ & $  0.273$ & $ +0.31\% $ & $0.063 $ \\ 
$T$ + $\delta$ &  $ -0.16\% $ & $ 0.0325  $ &  $-0.25\% $ & $ 0.0506 $ & $ -0.14 \% $ & $ 0.029$ & $   + 1.04  \% $ & $      0.208 $ \\ 
$T$ - $\delta$ &  $-0.18 \% $ & $ 0.0362 $ &  $ -0.28\% $ & $  0.056 $ & $   -0.16 \% $ & $   0.032$ & $     -0.01  \% $ & $  0.202 $ \\ 

$I_0$ + $\delta$ &  $ +  0.38\% $ & $0.076   $ &  $+ 1.72 \% $ & $ 0.345 $ & $ -0.27 \% $ & $  0.054$ & $  +2.97 \% $ & $   0.59 $ \\ 
$I_0$ - $\delta$ &  $   -0.74 \% $ & $ 0.149 $ &  $ -2.26 \% $ & $  0.452 $ & $   -0.059 \% $ & $ 0.012$ & $ -0.93 \% $ & $ 0.185 $ \\ 
\hline
\end{tabular}
\vskip -0.1cm
\caption{Sensitivity study with $\delta= 5\%$ of the parameters values in the reference set. In each row  we display  the volumes in $cm^2$ of the different components and the percentage variations of volume with reference to the set of values.}\label{tab1}  
\end{table}

Studying  Table \ref{tab1},  we can observe that some parameters like $\{ k_N, \epsilon, M \}$ appear to have a small influence on the main output of the model (the biofilms volume). As a consequence,  we can introduce some small variations on the parameter values without changing the simulation results. 
On the contrary, a parameter like $k_{B0}$ has a strong influence on the volume of biofilms. 
As expected by the model structure, the growth rate of cyanobacteria is a sensitive and pivotal parameter. With reference to the other rates, we can observe that $k_E$, the EPS growth rate, and $k_D$, the cyanobacteria death rate, have a moderate influence on the model results. 
Since the EPS is the bigger component in the biofilm, it is not surprising that this parameter plays an important role in the model. It could be extremely interesting then to  study with more attention this rate and to establish for this coefficient   the environmental parameters which can influence its variation, as done for $k_B$.  
Another parameter that should be studied more carefully  is $k_D$. Looking at Figure \ref{bio30},  we remark that the distribution of the dead component is proportional to the cyanobacteria distribution. In order to have a more realistic description of these rates, new in vitro experiments  are needed since, to our knowledge, data are still lacking.
The parameter $\gamma$ which represents the propagation of the front velocity, mildly influences the model outputs. Looking at the environmental parameters, i.e. the incident light intensity on the upper surface of the water $I_0$ and the temperature $T$, they influence directly the cyanobacteria growth since they are included in $k_B(I,T)$; then, it could be useful to insert these environmental conditions  in the growth of another important component of the biofilm, the extracellular matrix of polymeric substances.

\subsection{Robustness}
\begin{table}[ht!]
\centering
{\begin{tabular}{|l|l|l|} 
\hline
Parameter & Value in the reference set & Range of values  \\[3pt] \hline

$M$  &  $1\cdot 10^{-8}$ & $[-99.99 \%; + 9.999\cdot10^8 \%]$  \\ 
$k_N$  & $1\cdot  10^{-8}$ & $[-99.99 \%; +5.59\cdot10^6 \%]$  \\ 
$\epsilon$  &  $1\cdot 10^{-7}$ & $[-99.99 \%; +29400\%]$ \\
$\gamma$  & $5\cdot 10^{-16}$ & $[-99.99 \%; +24600 \%]$ \\
$k_D$ & $0.2\cdot 10^{-6}$ & $[-99.99 \%; +1900\%]$ \\
$k_E$  & $12\cdot 10^{-6}$ & $[-95.65 \%; + 1300 \%]$  \\ 
$k_{B0}$  &  $8\cdot 10^{-6}$ & $[-60 \% ; + 452.5\%]$ \\ 
$T$ & $30$ & $[-75\%; +300\%]$ \\
$I_0$ & $0.01$ & $[-75\%; +500\%]$ \\ \hline
\end{tabular}}
\vskip -0.1cm
\caption{Robustness study. Range of values in percentage for each parameter where the model still meets the criteria described in the section.}
\label{tab2}
\end{table}
In this subsection, our aim is to detect the ranges of parameter values which guarantee a quite stable behavior of the whole system.
By the robustness study we want to check if the model with different values of parameters would meet the
following rules: the final volume of cyanobacteria and dead cells after 30 days has to be lower than $1\,cm^2$ and higher than $0.01\,cm^2$, while the EPS volume has to be lower than $2\, cm^2$ and higher than $0.02\,cm^2$.

Hence,  we change the value of one parameter keeping all the other parameter values unchanged. Then we solve numerically the equations and check if the conditions are still satisfied. By the values obtained, reported in Table \ref{tab2}, we can sort the parameters in three classes:
\begin{itemize}
\item the parameters that have little influence on the model outputs ($M$, $k_N$);
\item the parameters that have a moderate influence on the model outputs ($\gamma$, $\epsilon$, $k_D$);
\item the parameters that have a strong influence on the model outputs ($k_E$, $k_{B0}$, $I_0$, $T$).
\end{itemize}
We can notice that, even if we highly change the values of some parameters ($M$, $k_N$, $\gamma$, $\epsilon$, $k_D$, and $k_E$), we obtain quite stable results. Therefore,  they affect mainly the related cells but
do not influence the whole process.

On the contrary, the interval ranges found for the cyanobacteria birth rate, temperature and light intensity are smaller and their modification brings to significant changes in the simulation results. It is interesting to notice that both sensitivity and robustness analyses indicate the same parameters as pivotal in the system dynamics.

Now that we have constructed a control on the parameters we use, we can perform some numerical simulations of the biofilm growth with the help of our model.

\section{Numerical simulations}

Numerical schemes used to solve system \eqref{systemfinal} in the two and three-dimensional cases are based on a finite differences
method in space and an implicit-explicit method in time. 
Our model presents two important differences with respect to a usual hyperbolic system. First, since we are considering a
multiphase fluid, it is difficult to deal with regions where one of the phases may vanish.
We solve this problem of vanishing phases by using an implicit-explicit scheme
in the approximation of the source terms. 
The second problem arises from the fact that our system is supplemented with a constraint term
due to the mass conservation, which implies that the average hydrodynamic velocity
of the mixture is divergence free. This constraint is needed to compute the hydrostatic
pressure. To enforce the divergence free constraint, we use a fractional step approach
similar to the Chorin-Temam projection scheme for the
Navier-Stokes equations, with a quite accurate reconstruction of the pressure term.
Details regarding the numerical scheme can be found in \cite{cdnr}.
 
\subsection{Long time simulations}
We perform a long time simulation ($10000$ days) in the two dimensional case to study the behavior of solutions to the the full system \eqref{systemfinal}. 
In Section 3 we have shown that for any initial condition, where $B_0>0$ in some points of the domain, the solutions to the system tend to the stationary points $\bar{B}$, $\bar{E}$, $\bar{D}$ and $\bar{L}$ defined at equation \eqref{stat2}.
To control this result,  we perform a simulation using a constant incident light and neglecting its absorption by water, such that $k_B$ has a constant value in the whole domain.  We consider as initial data a distribution of $5$ small gaussian functions of cyanobacteria, with a total volume of $2.0356 \cdot 10^{-4}\, (cm^2)$ while EPS and dead cells are  equal to zero. The domain is the square $\Omega = [0,L]\times [0,L]$ where $L=5\, cm$.
Results are reported in Table \ref{tab4}, where we list the maximal values  in space reached by $B$, $E$, $D$, respectively for each time instant reported on the last column on the right.
\begin{table}[ht!]
\centering
      
\begin{tabular}{|l|l|l|l|}
\hline
Cyanobact. (max) & EPS (max) & Dead C. (max)  & Time (days) \\
\hline
0.3587 & 0.5483 & 0.0067  & 10  \\
0.3706 & 0.6162 &0.0739  & 60  \\
0.3707 & 0.6179 &0.1157   & 100  \\
0.2101 & 0.5981 &   0.3426  & 500  \\
0.1411 & 0.4365 &0.4621  & 1000  \\
0.1134 & 0.3423 &  0.5290 & 2000 \\ 
0.1084 &0.3251 &  0.5416  & 5000 \\ 
0.1083 $*$ &0.325 $*$ &  0.5417$*$ & 10000 \\
\hline
\end{tabular}
\vskip -0.1cm
\caption{Maxima with respect to  time.}
\label{tab4} 
\end{table}

Let us observe that  in this case, the stationary values are reached after $10000$ days. Here the symbol $*$ added after the numerical value in the last row means that the maximum and the minimum values coincide, and they correspond to the stationary values of homogeneous solutions, see eq. \eqref{stat2}: $\bar{B} =    0.1083$, $\bar{E} = 0.325$ and $\bar{D} = 0.5417$.




\subsection{Biofilm growth}

\begin{figure}[!ht]
\centering
        \includegraphics[width=9.5cm, height=5cm]{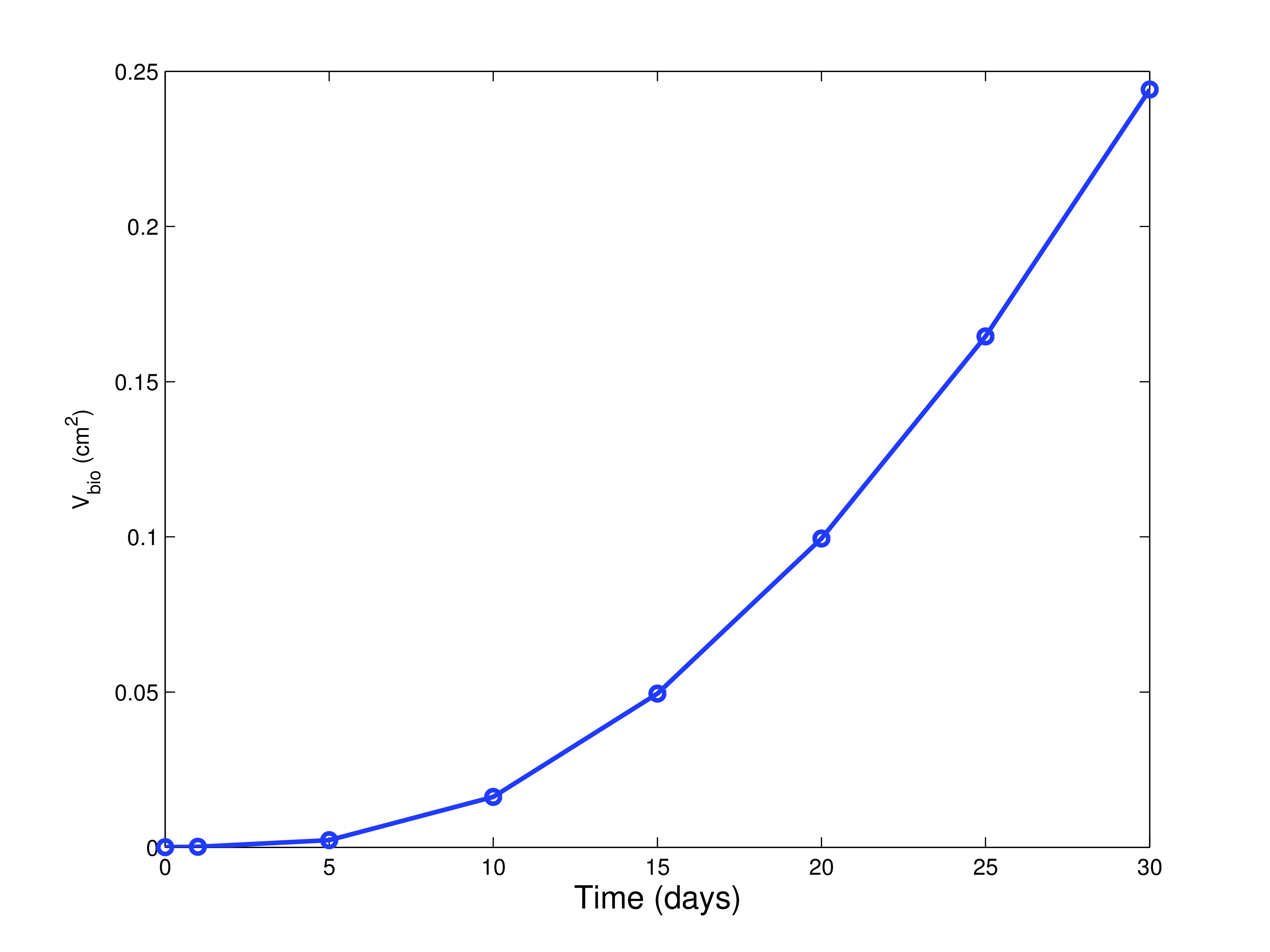}
\vskip-0.2cm
  \caption{\footnotesize{Biofilm volume growth during the first $30$ days (active growth phase). The growth has an exponential behavior at the beginning and after 15 day we can observe a linear behavior.}}
    \label{vbiot}
\end{figure}

Initially, we are interested in observing the biofilm growth in the first $30$ days. Thus, we simulate the growth of the biofilm volume components during the first $30$ days (active growth). We consider as an  initial condition a gaussian distribution of cyanobacteria situated  in the center of the domain like \eqref{initialdata2d} with an initial total volume $V_{B0} = 4.071 \cdot 10^{-5} \, (cm^2) $. Since we consider a colony of cyanobacteria, the values of the other biomasses are initially zero.
In Table \ref{tabgro} we present the variation of volumes as a function of  time during the first $30$ days. In Figure \ref{vbiot}  we can observe that in the active phase, at the beginning,  the biofilm volume has  an exponential growth and, after 15 days,  due to the limiting growth factors present in the model,  the growth becomes linear.
\begin{table}[ht!]
\centering
{\begin{tabular}{|l|l|l|l|l|} \hline
$Time\, (days)$ & $V_{bio}\, (cm^2)$ & $V_{B} \, (cm^2)$ & $V_E \, (cm^2)$ & $V_D \, (cm^2)$ \\ \hline
$1$ & $1.3834 \cdot 10^{-4}$ & $ 0.7944 \cdot 10^{-4}$ & $0.5864 \cdot 10^{-4}$ & $2.4815 \cdot 10^{-7}$ \\
$5$  & $ 0.0023 $ & $0.00092$ & $0.0013$ & $6.4863 \cdot 10^{-6}$ \\
$10$ & $0.0162$ & $0.0061$ & $0.01$ & $7.2077 \cdot 10^{-5}$ \\ 
$15$ & $0.0495$ & $ 0.0176$ & $0.0316$ & $3.1989 \cdot 10^{-4}$ \\
$20$ & $0.0995$ & $0.0334$ & $0.0653$ & $8.6315\cdot 10^{-4}$ \\
$25$ & $0.1646$ & $ 0.0525$ & $0.1104$ & $0.0018$ \\
$30$ & $0.2441$ & $0.0745$ & $0.1665$ & $0.0031$ \\ \hline
\end{tabular}}
\vskip-0.1cm
\caption{Volume growth as a function of  time of biofilm volume $V_{Bio}$, cyanobacteria volume $V_B$, EPS volume $V_E$ and dead cells volume $V_D$.}
\label{tabgro}   
\end{table} 

Since the sensitivity and robustness analyses indicate the incident light intensity and the temperature as quite important parameters in the model dynamics, it is interesting to observe the model response under different values of these two environmental factors.
Assuming to have an incident light $I_0$ and an optimal light $I_{opt} = 0.01 \, (\mu mol \, cm^{-2}\,sec^{-1})$, we use a varying incident light intensity $I_0$, with a range from $0.01 \cdot I_0$ to $10 \cdot I_0$. Moreover, we assume that the light is absorbed by water and by cells with the coefficient values reported in Table \ref{partab}. 
Results of simulations are listed in Table \ref{tab3}, where we can observe the influence of incident light $I_0$ when it moves away from $I_{opt}$ in an increasing or decreasing verse. In this manner, the variations of incident light influence the growth of biofilm and,  in particular, cyanobacteria and EPS are more sensible to light variations, as expected from the model. 

\begin{figure}[!ht]
    \centering
        \includegraphics[width=7cm, height=4cm]{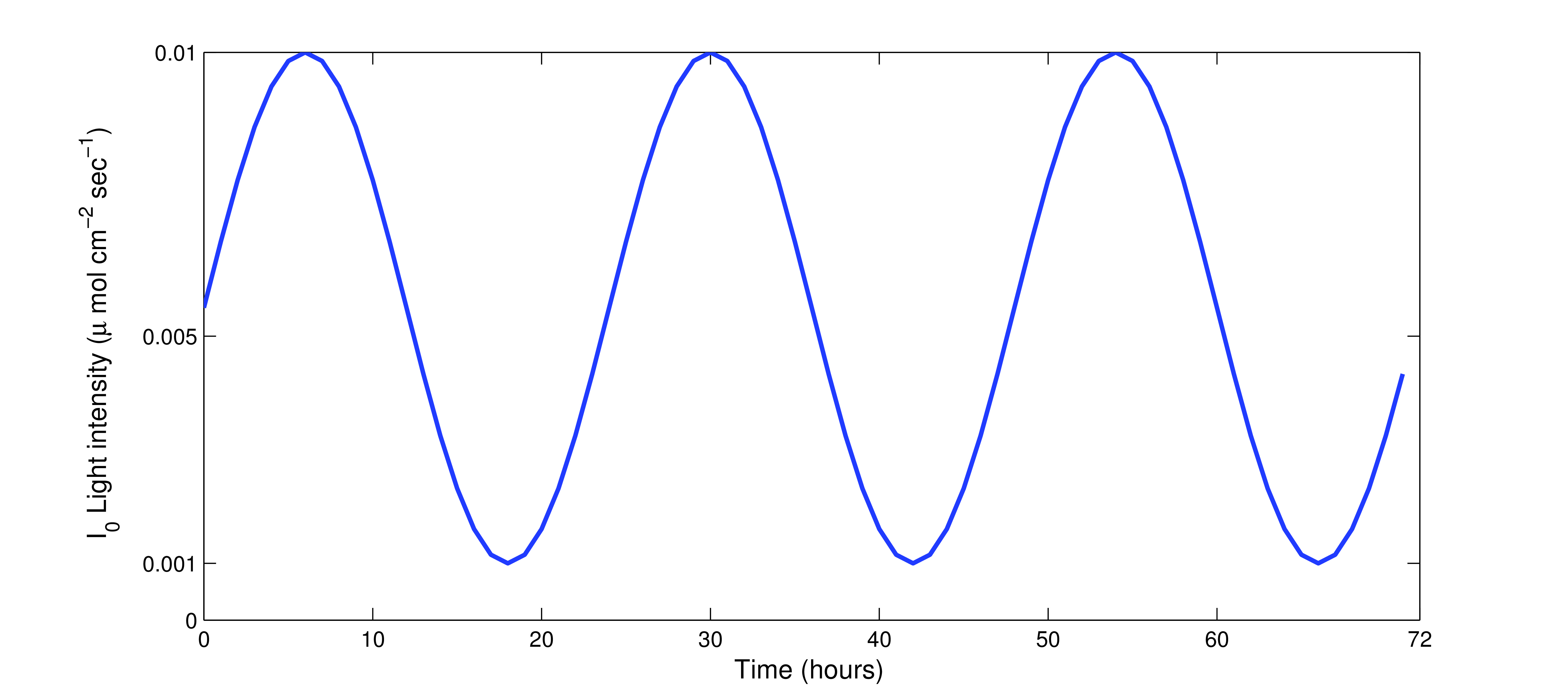}    \includegraphics[width=7cm, height=4cm]{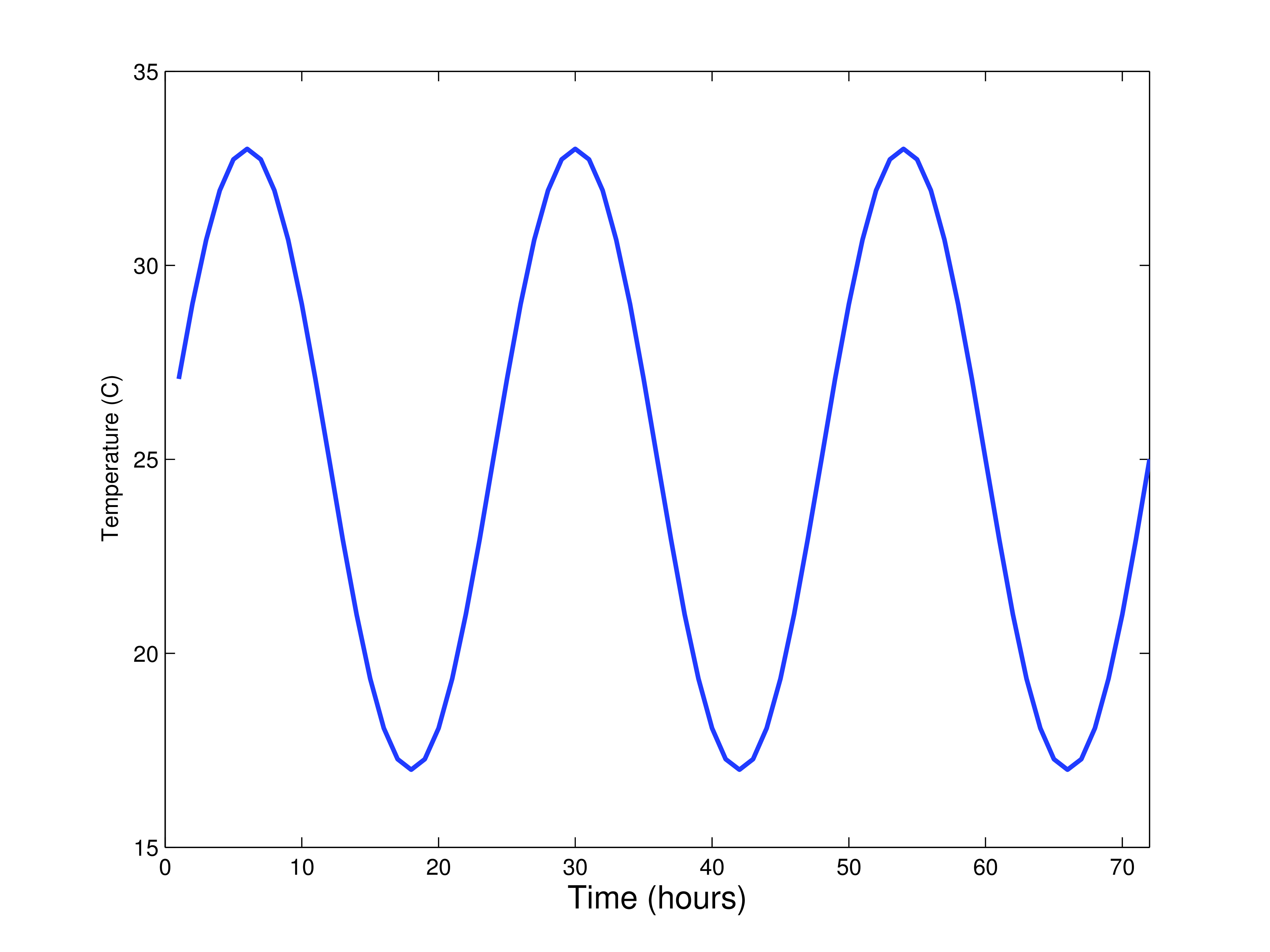}
    \caption{\footnotesize{On the right the variable incident light intensity $I_0$ and on the left the variable Temperature $T$ reproducing a day-night cycle.}}
    \label{I0}
\end{figure}

In the last row of  Table \ref{tab3}, in order to simulate a day-night cycle, we consider a sinusoidal variation of light with a period of $24$ hours. Figure \ref{I0} represents the sinusoidal function where the amplitude of oscillations has a maximum of incident light $I_0=I_{opt}$ during the day and a minimum  equal to $0.1 \cdot I_0$ during the night.
\begin{table}[ht!]
\centering
{\begin{tabular}{|l|l|l|l|l|} \hline
Light intensity & $V_{bio}\, (cm^2)$ & $V_{B} \, (cm^2)$ & $V_E \, (cm^2)$ & $V_D \, (cm^2)$ \\ \hline
$I_0/100$ & $ 0.0015 $ & $ 5.1682\cdot10^{-5}$ & $0.0014$ & $6.995 \cdot 10^{-6}$ \\
$I_0/50$ & $0.0018 $ & $ 7.7003\cdot10^{-5}$ & $0.0018$ & $8.6052 \cdot 10^{-6}$ \\
$I_0/20$ & $ 0.0036 $ & $2.3251\cdot10^{-4}$ & $0.0033$ & $1.6606\cdot 10^{-5}$ \\
$I_0/10$ & $0.0101$ & $0.001$ & $0.009$ & $4.8503 \cdot 10^{-5}$ \\ 
$0.4 \cdot I_0$ & $0.1231$ & $0.0258$ & $0.0963$ & $ 0.0011$ \\
$0.6 \cdot I_0$ & $0.1912$ & $0.0484$ & $0.1408$ & $0.002$ \\ 
$0.8 \cdot I_0$ & $0.2297$ & $0.0652$ & $0.1618$ & $0.0027$ \\
$I_0$ & $0.2441$ & $0.0745$ & $0.1665$ & $0.0031$ \\ 
$1.2 \cdot I_0$ & $0.2432$ & $0.0777$ & $0.1622$ & $0.0033$ \\
$1.5 \cdot I_0$ & $0.2271$ & $ 0.0754$ & $0.1486$ & $0.0031$ \\
$2.0 \cdot I_0$ & $0.1892$ & $0.0639$ & $0.1228$ & $ 0.0026$ \\
$5.0 \cdot I_0$ & $0.061$ & $0.017$ & $ 0.0434$ & $5.6982 \cdot 10^{-4}$ \\
$10 \cdot I_0$ & $ 0.0156$ & $0.0026$ & $0.013$ & $8.5515 \cdot 10^{-5}$ \\ \hline
$I_0$ sinusoidal & $ 0.1362$ & $ 0.0312$ & $ 0.1037$ & $ 0.0013$ \\ \hline
\end{tabular}}
\vskip-0.1 cm
\caption{Volume growth in $30$ days with different values of the incident light intensity $I_0$.}
\label{tab3}
\end{table}

We perform a similar study to determine  the temperature influence on the biofilm growth.
It is known from experiments that there is an optimal temperature for cyanobacteria development and, according to \cite{dip09} and \cite{alb08}, we take $T_{opt}=30^o C$.
We use then a varying temperature $T$, with a range from $0.01 \cdot T$ to $10 \cdot T$. 
Results of simulations are listed in Table \ref{tab3bis}, where we can observe the influence of temperature $T$ when it moves away from $T_{opt}$ in an increasing or decreasing  verse.
In the last row of Table \ref{tab3bis}, we consider a sinusoidal variation of temperature with a period of $24$ hours to simulate a day-night cycle. As shown in Figure \ref{I0},  the amplitude of oscillations is $T_{min}=17^o C $ which corresponds to the minimal temperature during the night and a maximal temperature $T_{max}=33^o C$, reached during the day.
\begin{table}[ht!]
\centering
{\begin{tabular}{|l|l|l|l|l|} \hline
Temperature & $V_{bio}\, (cm^2)$ & $V_{B} \, (cm^2)$ & $V_E \, (cm^2)$ & $V_D \, (cm^2)$ \\ \hline
$T/100$ & $ 0.0015 $ & $ 5.2789\cdot10^{-5}$ & $0.0014$ & $7.0534 \cdot 10^{-6}$ \\
$T/50$ & $0.0019 $ & $ 8.0606\cdot10^{-5}$ & $0.0018$ & $8.772 \cdot 10^{-6}$ \\
$T/20$ & $  0.0039 $ & $  2.6662\cdot10^{-4}$ & $0.0036$ & $1.7796\cdot 10^{-5}$ \\
$T/10$ & $ 0.0118$ & $0.0013$ & $ 0.0104$ & $  5.8265 \cdot 10^{-5}$ \\ 
$0.4 \cdot T$ & $0.1363$ & $0.0338$ & $0.1012$ & $ 0.0014$ \\
$0.6 \cdot T$ & $0.2021$ & $0.0576$ & $0.1421$ & $0.0024$ \\ 
$0.8 \cdot T$ & $0.2353$ & $0.0709$ & $ 0.1615$ & $0.003$ \\
$T$ & $0.2441$ & $0.0745$ & $0.1665$ & $0.0031$ \\ 
$1.2 \cdot T$ & $ 0.2382$ & $0.072$ & $0.1631$ & $0.003$ \\
$1.5 \cdot T$ & $0.2165$ & $ 0.0633$ & $0.1506$ & $0.0031$ \\
$2.0 \cdot T$ & $0.1734$ & $ 0.0468$ & $0.1248$ & $ 0.0019$ \\
$5.0 \cdot T$ & $0.0467$ & $0.0082$ & $0.0382$ & $3.1116\cdot 10^{-4}$ \\
$10 \cdot T $ & $ 0.0118$ & $0.0013$ & $   0.0104$ & $5.8265 \cdot 10^{-5}$ \\ \hline
$T$ sinusoidal & $0.228$ & $ 0.0678$ & $0.1573$ & $0.0028$ \\ \hline
\end{tabular}}
\vskip-0.1cm
\caption{Volume growth in $30$ days with different values of temperature $T$.}
\label{tab3bis}
\end{table}

\subsection{Simulations in  the two-dimensional case}
We reproduce numerically the active growth of a phototropic biofilm for a period of $30$ days taking into account the indications given  in \cite{dip09,alb08,zip07} that we have already used in the calibration and estimate of coefficients.
In particular, these laboratory experiments are performed using "Truelight lamps" (Auralight, Sweden), with a daily-cycle given by $16$ light hours and $8$ hours of night. We reproduce this setting of illumination adopting a light-night cycle ($16-8$) and assuming a constant temperature $T = T_{opt} = 30^oC$. 
We take as an  initial condition a distribution of $5$ small gaussian functions of cyanobacteria placed on the bottom of the square domain $\Omega=[0,5]\times[0,5]$ $cm^2$ with a total volume of $V_{B_0}=2.0356 \cdot 10^{-4}\, (cm^2)$, while initial EPS and dead cells are equal to zero.
In this case space variables account for width and height, considering that  all functions are constant in length.
Under these conditions, we simulate a $30$ days growth of a biofilm, using the parameter values listed in Table \ref{partab}.  

\begin{figure}[!ht]
    \centering
        \includegraphics[width=5cm]{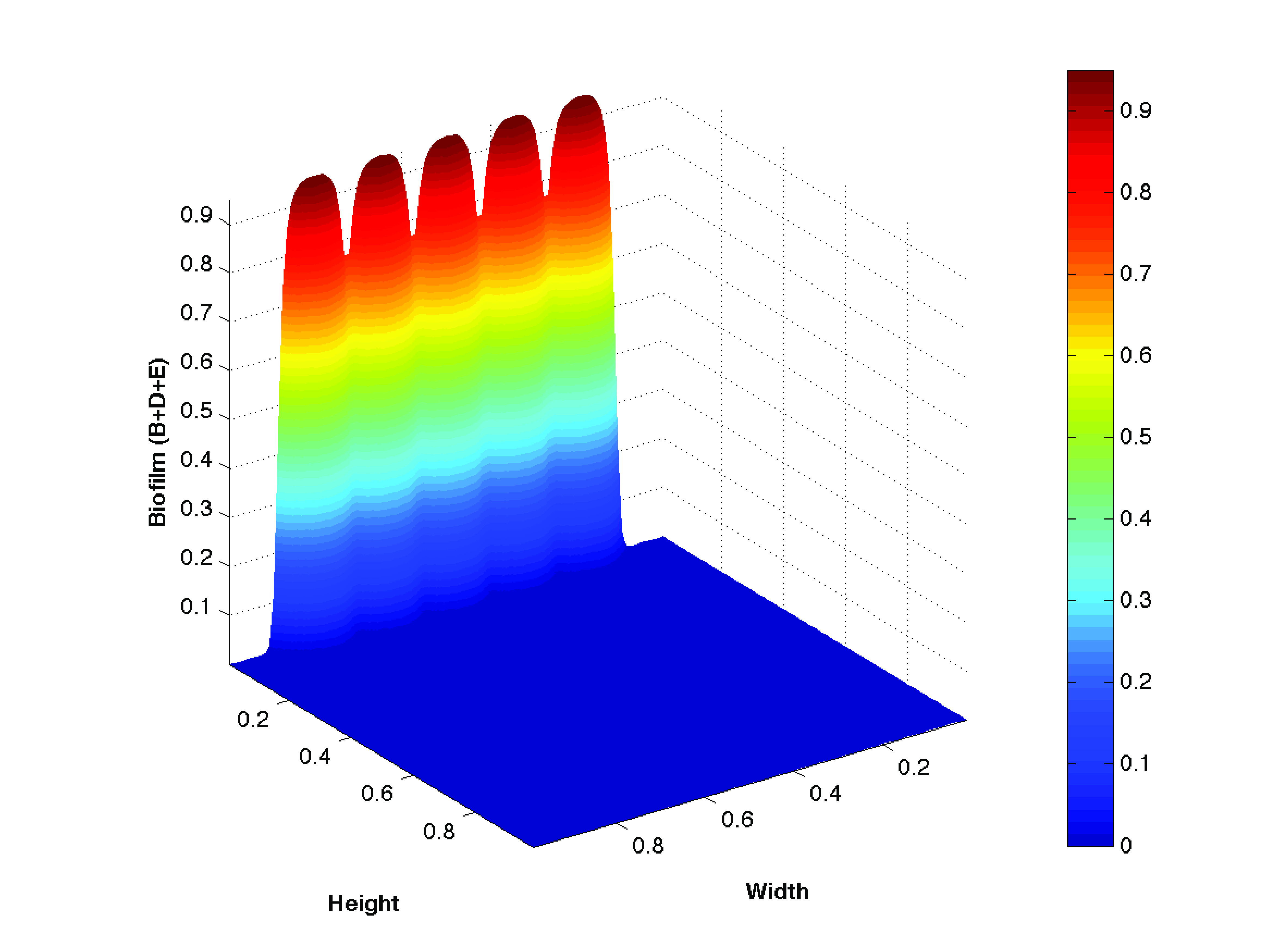}
        \includegraphics[width=5cm]{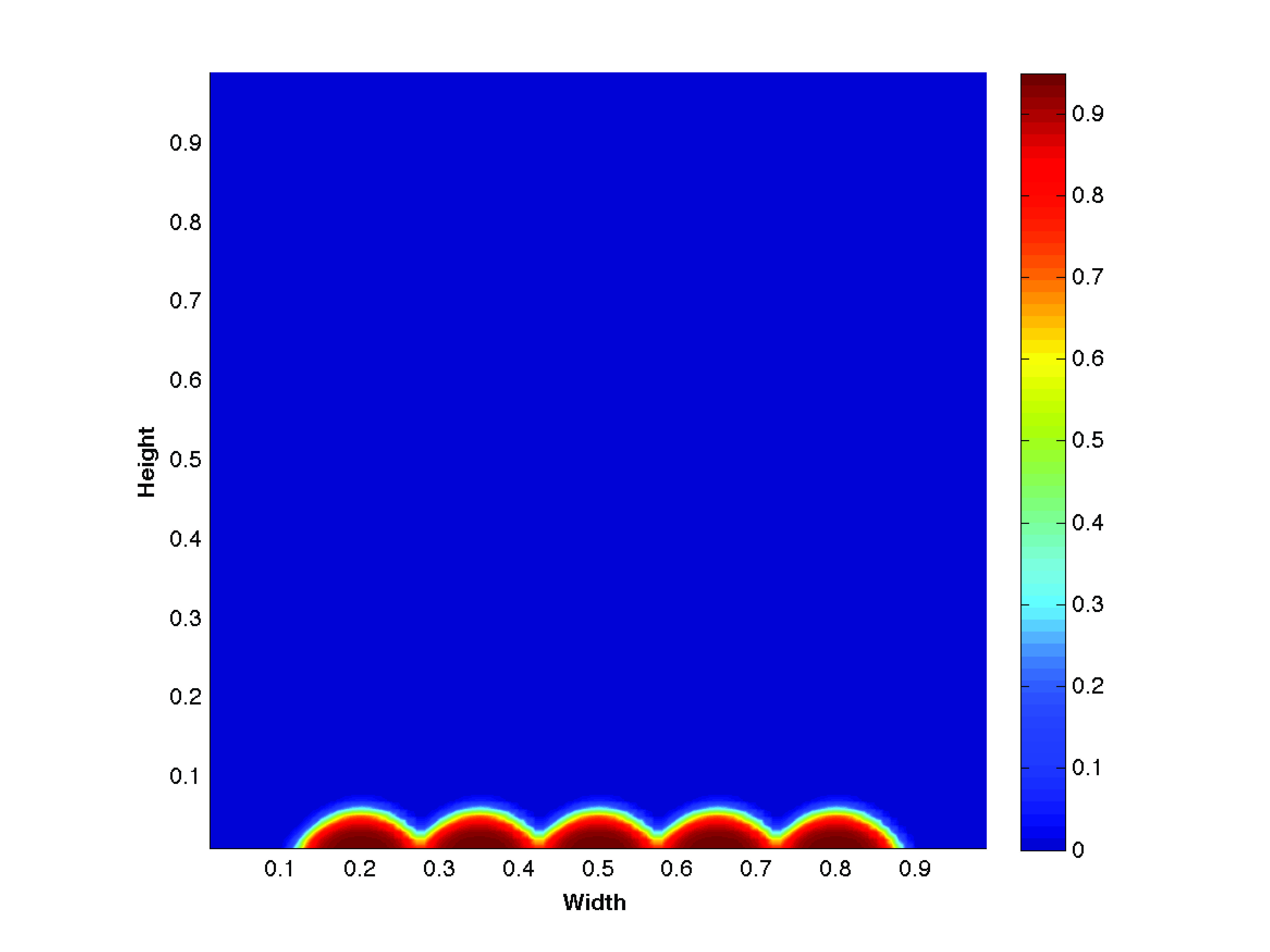} \includegraphics[width=5cm]{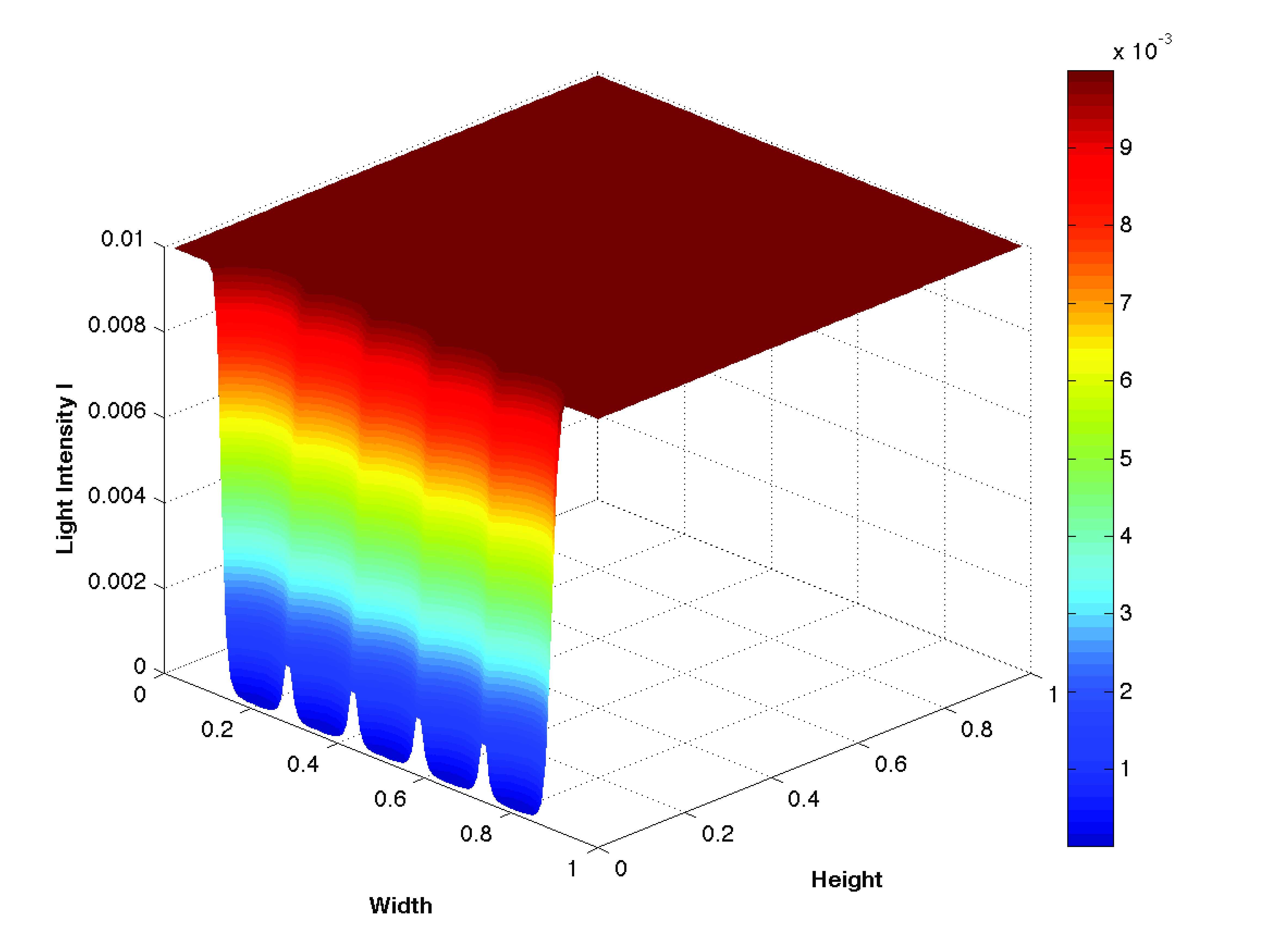}
    \caption{\footnotesize{Biofilm growth after $30$ days in a domain $5\times5$ ($cm^2$). On the left a three-dimensional vision of biofilm. In the middle a point of view from side of the biofilm in the 2-dimensional domain and on the right the light distribution throughout the domain.}}
    \label{bio_sim}
\end{figure}

A plot of the results is displayed in Figure \ref{bio_sim}: on the left  we observe the formation of an homogeneous layer of components by a quick aggregation, in the middle is presented a view from the side of the biofilm as a function of width and height and on the right we show the light intensity distribution after $30$ days (after the last switch off). Since the biomasses absorb much more photons than the water molecules, the boundary between the light distribution through  the liquid and the biofilm, where the light is damped, is quite clear. The total final volume of the whole biofilm is $V_{Bio} = 0.5980 \, (cm^2)$. The maximum thickness of this simulation is about $2-3$ millimeters; if we guess to have an homogeneous distribution of the biofilm volume, considering the basis of $5$ $cm$, we obtain an average thickness of $0.1196$ $cm$. This result  is in agreement with the experiments performed by biologists on cyanobacteria biofilms under the same conditions of light and temperature, see  \cite{dip09}, \cite{alb08} and \cite{zip07}.  


\subsection{Simulation in  the three-dimensional case}
\begin{figure}[!ht]
    \centering
        \includegraphics[width=4.8cm]{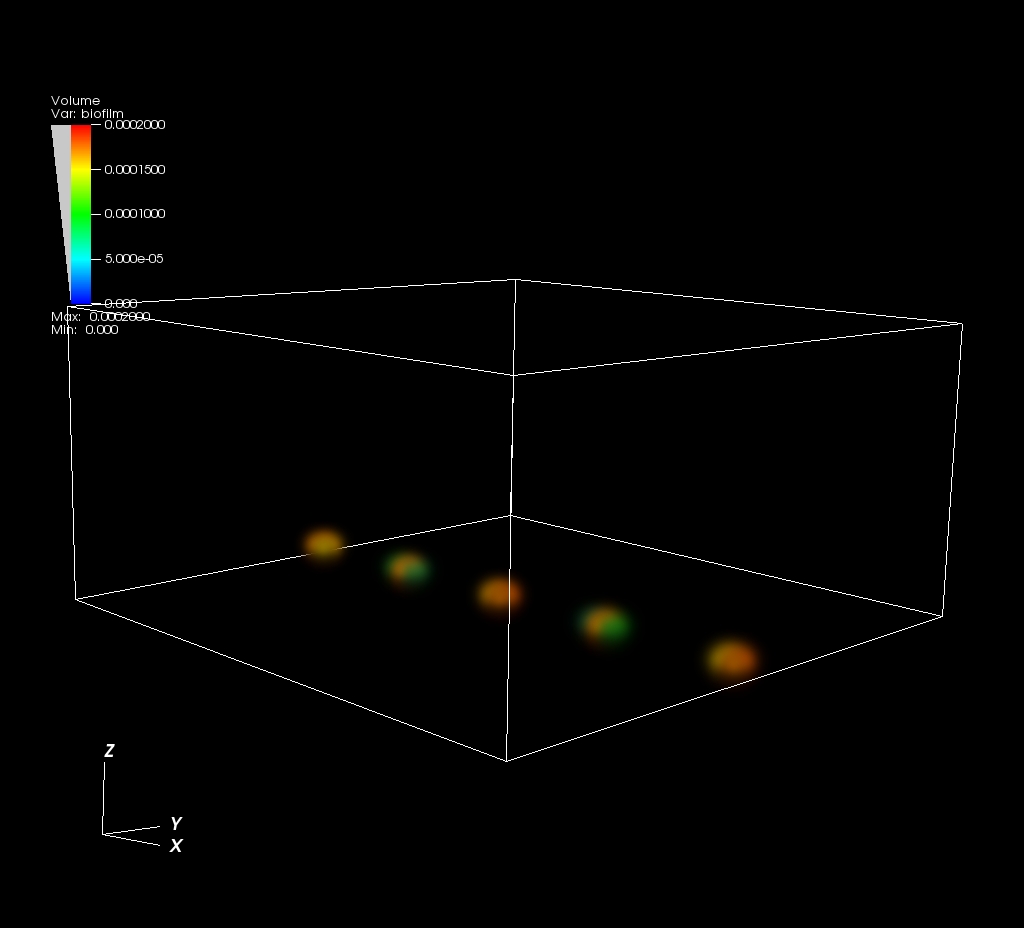} \includegraphics[width=4.8cm]{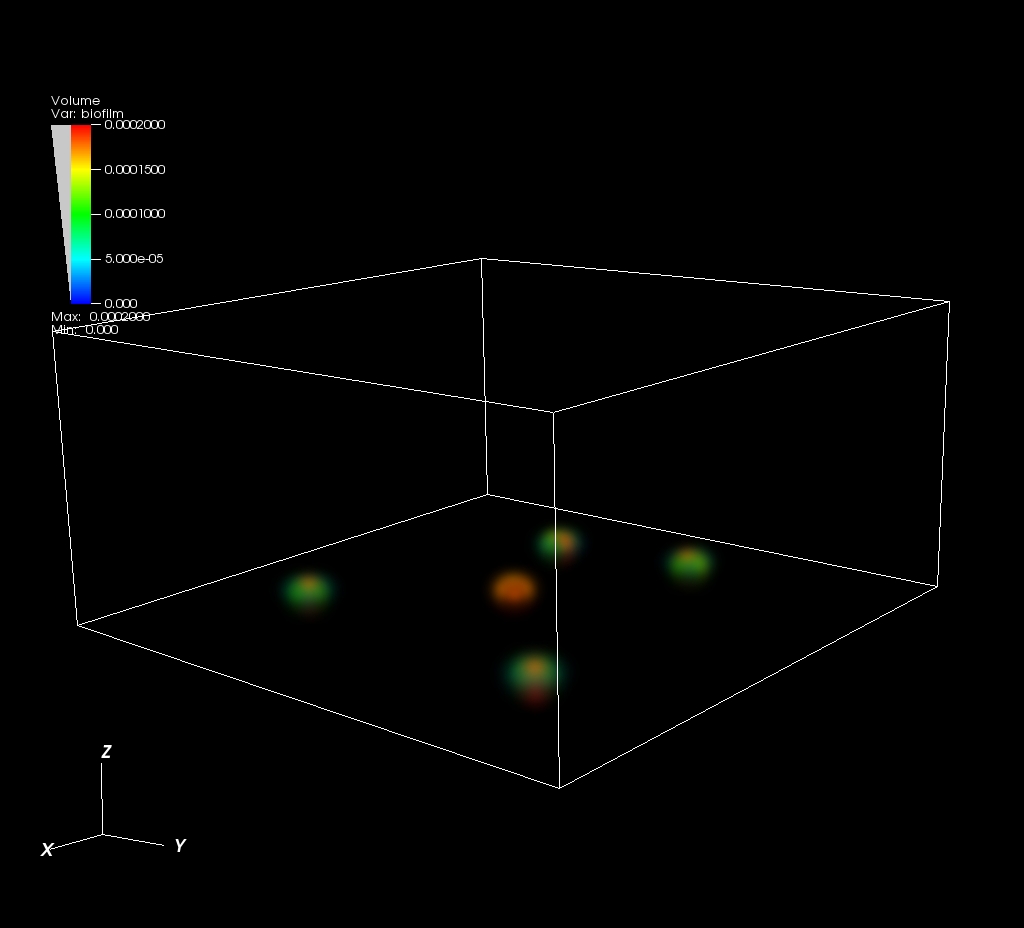}\\ 
        
        \includegraphics[width=4.8cm]{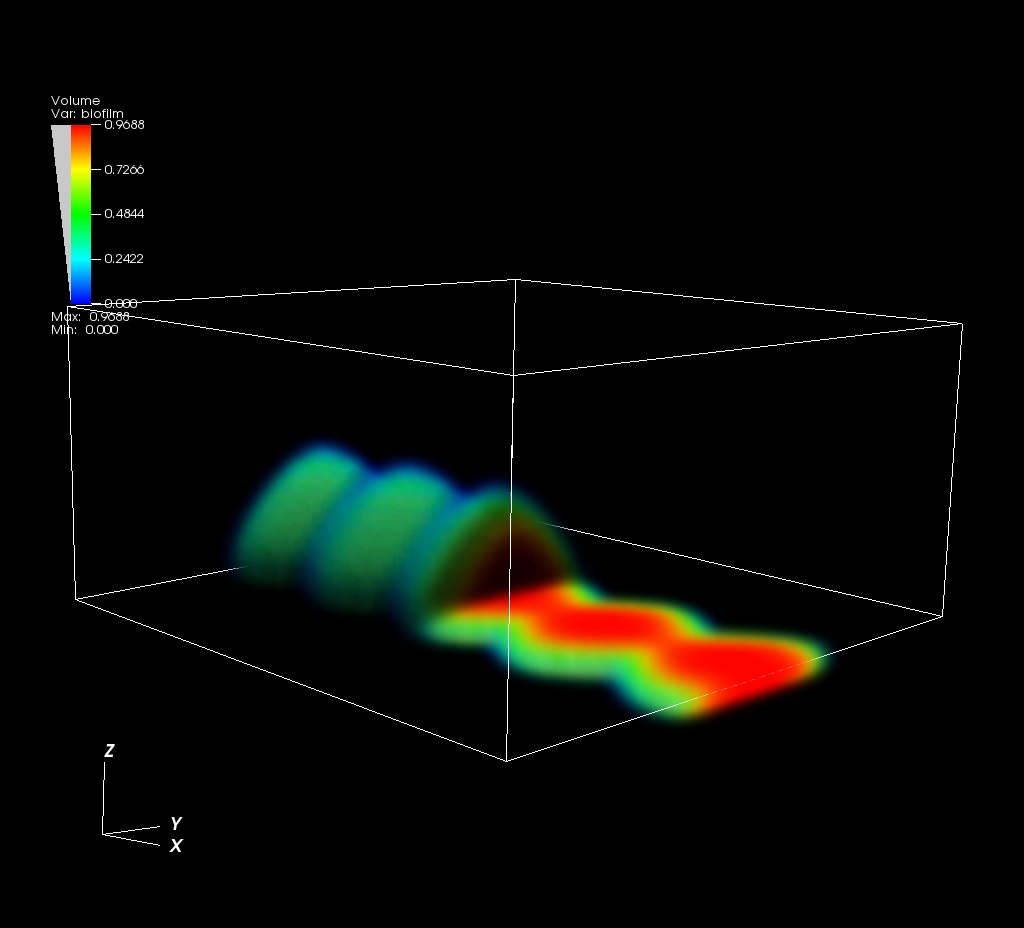} \includegraphics[width=4.8cm]{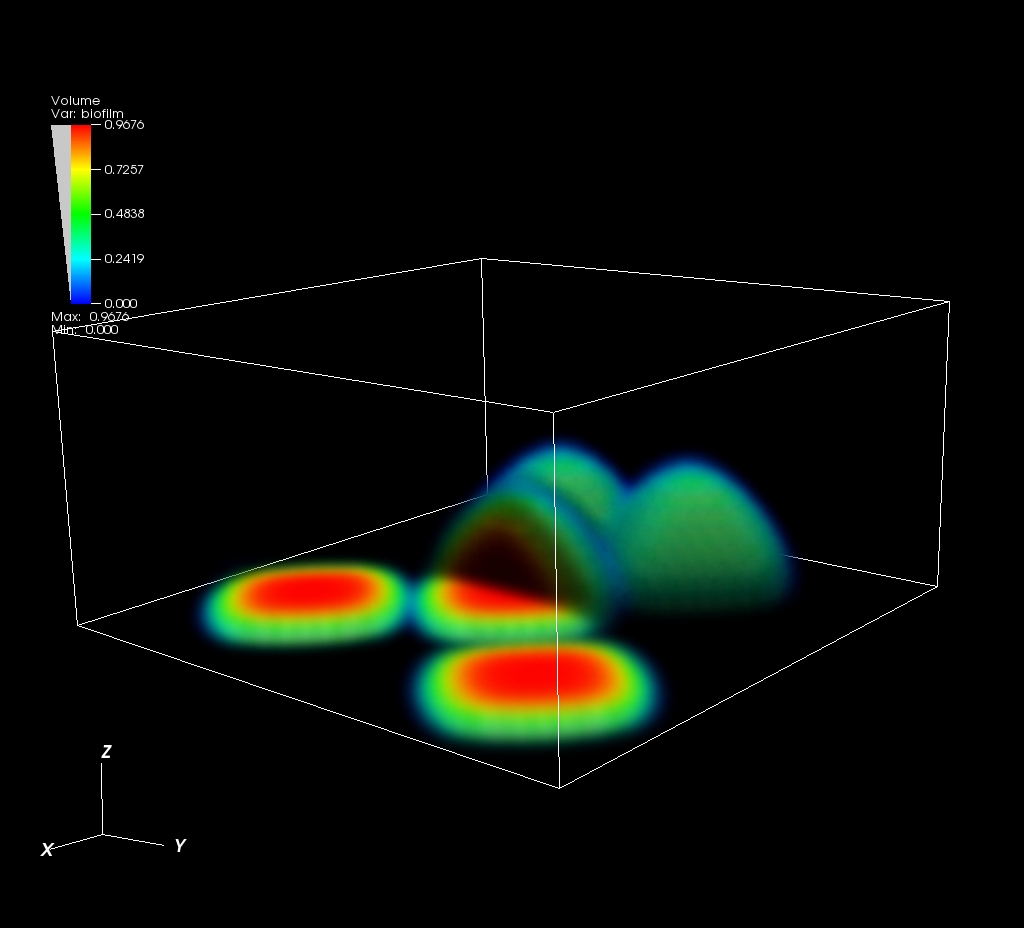}\\
        
        \includegraphics[width=4.8cm]{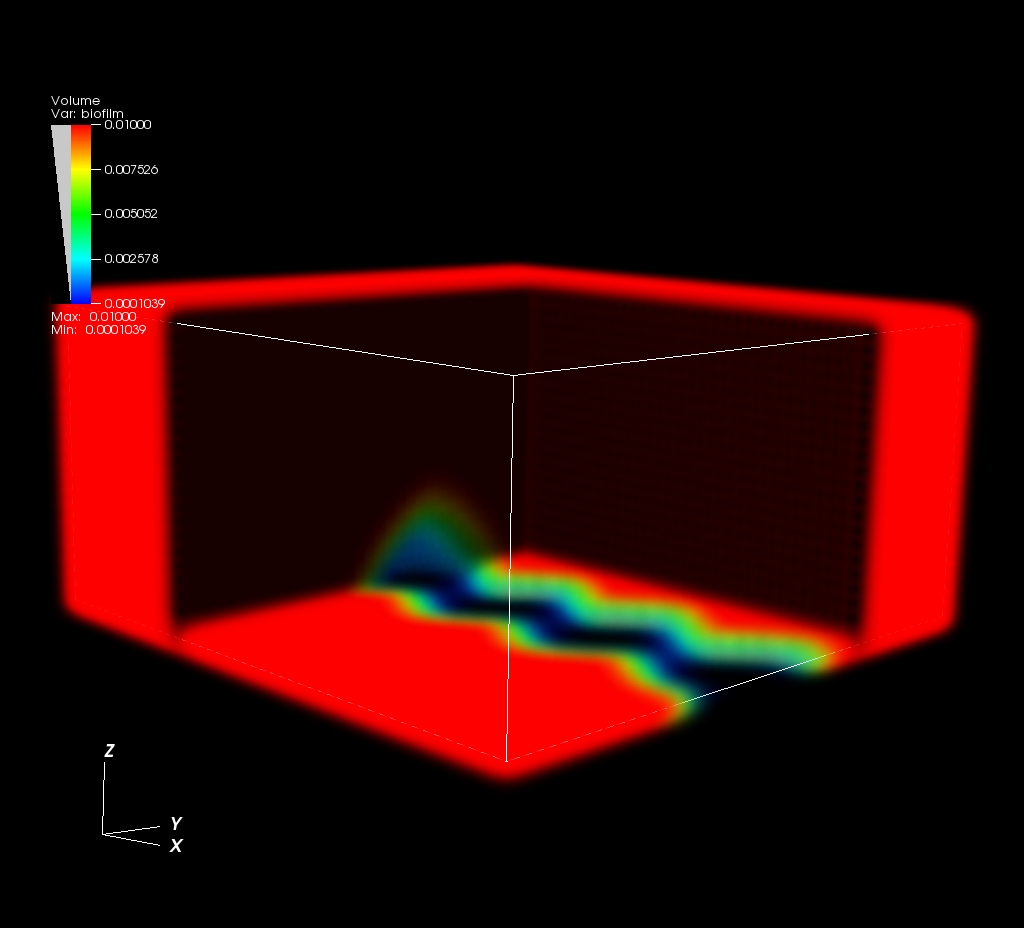} \includegraphics[width=4.8cm]{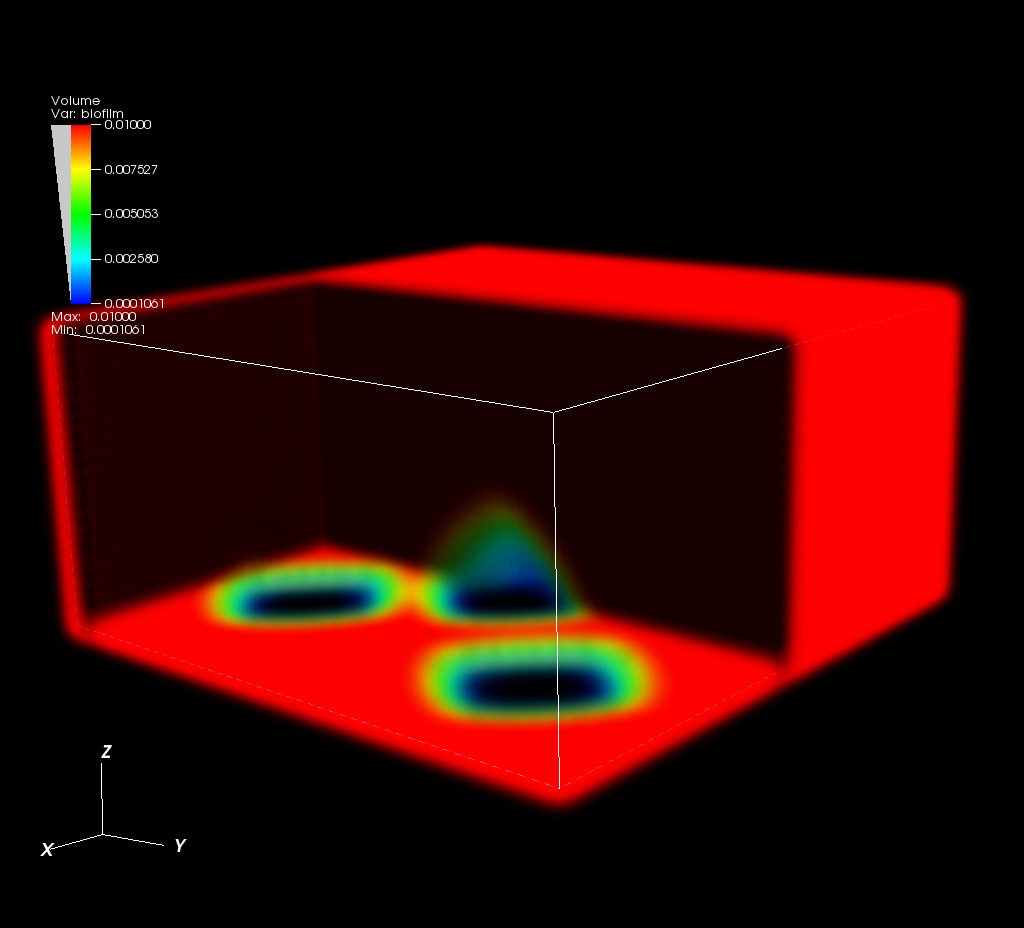}
    \caption{\footnotesize{Biofilm growth after $30$ days in a domain $1\times  1 \times 0.5$ $(cm^3)$ with two different initial conditions. On the top the initial conditions, in the middle the final evolutions of biofilm and on the bottom the light distribution throughout the domain. Sections are made to emphasize the internal structure of the biofilm and of the distribution of light.}}
    \label{bio_3D}
\end{figure}

In this last section we present the three dimensional simulations of active growth of a phototropic biofilm for a period of $30$ days. 
We use the values of parameters calibrated in the previous sections and listed in Table \ref{partab}.
In our simulations the three dimensional domain is $ [0,1]\times[0,1]\times[0, 0.5]\, (cm^3)$ and
the initial condition for the cyanobacteria volume fraction $B$ is a sum of $5$ Heaviside functions whose
amplitude is  of the order of the cell dimensions $h_x = h_y = h_z = 0.02$ $(cm)$,  with a total volume of $V_{B_0}=8\cdot 10^{-9}\, (cm^3)$ while the other components are initially equal to zero.
As reported in Figure \ref{bio_3D},  we consider two different initial conditions for cyanobacteria: on the left,  the five aggregates are aligned  in the center of the domain, while, on the right, the five aggregates are randomly distributed.

Moreover,  we can observe in the last line of Figure \ref{bio_3D} how the light intensity $I(\textbf{x}, t )$ is attenuated by the biomasses. 
Even if it  is not possible to compare our numerical results with experimental data in a fully quantitative way,  we can observe that also in the three dimensional case the order of height of biofilm we find is in agreement with experiments.  
As a matter of fact,  the total final volume of the whole biofilm at 30 days is $V_{Bio} = 0.0227 \, (cm^3)$. We can see  in Figure \ref{bio_3D} that  the maximum of the  thickness is about $2-3$ millimeters.  Considering the simulation where the initial data are aligned  in the center of the domain,  we remark an homogeneous distribution of the biofilm on a layer of 0.2 $cm$ of width, obtaining then an average thickness of $0.1135$ $cm$.

%
%

\section{Conclusion}

This work had as a main goal of analyzing the response of our fluid dynamical  model of biofilms under different values of environmental factors, in particular the incident light. Some experimental papers on  biofilm development have been considered in order to estimate some parameters and our analysis was able to assess the sensitivity and the robusteness of the main parameters. Moreover, our numerical simulation showed a good agreement with the results of the linearized analysis. However, it should be important to set up some new devoted  experiments  to measure the thickness growth of a biofilm in time, its light absorption, its temperature and nutrient dependence with a good knowledge of  initial data in order to obtain a more accurate calibration of the model.

\section*{Acknowledgements.}

This work has been partially supported by  the ANR project  MONUMENTALG,  ANR-10-JCJC  0103.

\end{document}